\documentclass[preprint]{aastex}
\usepackage{graphicx,amsmath,color,natbib}

\newcommand{\K}{\,{\rm K}}
\newcommand{\Teff}{T_{\rm eff}}

\newcommand{\ang}{\,\mbox{\AA}}
\newcommand{\fmx}{f_{\rm{m}}}
\newcommand{\vlos}{v_{\rm{los}}}
\newcommand{\kms}{\,\rm{km\,s^{-1}}}
\newcommand{\kpc}{\,\rm{kpc}}
\newcommand{\pc}{\,\rm{pc}}

\newcommand{\vsun}{\mathbf{v}_{\odot}}

\newcommand{\masyr}{\,\rm{mas\,yr^{-1}}}

\newcommand{\nplatesuniq}{937}          % N unique plates
\newcommand{\nplatesredund}{77}         % N redundant plates
\newcommand{\nbhbbr}{733}               % N g<18 pass combo cut
\newcommand{\nbhbbrnotstringent}{172}   % N g<18 pass combo cut but not pass s
\newcommand{\nnotbryesstringent}{199}   % N g<18 not pass combo cut but pass s
\newcommand{\nbsbryesstringent}{133}    % N g<18 not combo, pass scolor, BS by
\newcommand{\nbhbfaint}{437}            % N g>18 pass scolor cut
\newcommand{\nbhb}{1170}                % N final BHB
\newcommand{\ncolorcut}{4515}           % N pass color cut
\newcommand{\ncolorcutbr}{2338}         % N g<18 pass color cut
\newcommand{\ncolorcutdup}{81}          % N duplicates among those passing col
\newcommand{\ndupfromredund}{410}       % N duplicates among redundant plates
\newcommand{\nduptotal}{491}            % ncolorcuptdup+ndupfromrefund
\newcommand{\nbrbhbbhb}{91}             % N g<18 doubly classified BHB
\newcommand{\nbrbhbnot}{13}             % N g<18 conflicting classification
\newcommand{\nbrnotnot}{131}            % N g<18 doubly classified non-BHB
\newcommand{\rankcorr}{0.263}           % rank corr (g, Delta v)
\newcommand{\rmsverror}{36.5}           % rms(Delta v)
\newcommand{\lastplatedate}{2003 Aug 6}

\begin{document}

\title{Blue horizontal branch stars in the Sloan Digital Sky Survey:\\
	I. Sample selection and structure in the Galactic halo}
\author{
Edwin~Sirko\altaffilmark{\ref{princeton}},
Jeremy~Goodman\altaffilmark{\ref{princeton}},
Gillian~R.~Knapp\altaffilmark{\ref{princeton}},
Jon~Brinkmann\altaffilmark{\ref{apo}},
{\v Z}eljko~Ivezi\'c\altaffilmark{\ref{princeton}},
Edwin~J.~Knerr\altaffilmark{\ref{princeton},\ref{minnesota}}
David~Schlegel\altaffilmark{\ref{princeton}},
Donald~P.~Schneider\altaffilmark{\ref{pennstate}},
Donald~G.~York\altaffilmark{\ref{chicago}}
}
\altaffiltext{1}{Princeton University Observatory, Princeton,
	NJ 08544\label{princeton}}
\altaffiltext{2}{Apache Point Observatory, P.O. Box 59, Sunspot, 
	NM 88349\label{apo}}
\altaffiltext{3}{Law School, University of Minnesota, 
	Minneapolis, MN 55455\label{minnesota}}
\altaffiltext{4}{Department of Astronomy and Astrophysics, 
	the Pennsylvania State University, University Park, 
	PA 16802\label{pennstate}}
\altaffiltext{5}{Department of Astronomy and Astrophysics, 
	University of Chicago, 5640 South Ellis Avenue, 
	Chicago, IL 60637\label{chicago}}

\begin{abstract}

We isolate samples of $\nbhbbr$ bright ($g < 18$) and $\nbhbfaint$
faint ($g > 18$) high-Galactic latitude blue horizontal branch stars 
with photometry and spectroscopy in the Sloan Digital Sky Survey
(SDSS).  Comparison of independent photometric and spectroscopic
selection criteria indicates that contamination from F and
blue-straggler stars is less than $10\%$ for bright stars ($g<18$) and
about $25\%$ 
for faint stars ($g>18$), and this is qualitatively
confirmed by proper motions based on the USNO-A catalog as first
epoch.  Analysis of repeated observations shows that the errors in
radial velocity are $\approx 26 \kms$.

A relation between absolute magnitude and color is established using
the horizontal branches of halo globular clusters observed by SDSS.
Bolometric corrections and colors are synthesized in the SDSS filters
from model spectra.  The redder stars agree well in absolute magitude
with accepted values for RR Lyrae stars.  The resulting photometric
distances are accurate to about $0.2$ magnitudes, with a median of
about 25 kpc.
Modest clumps in phase space exist and are consistent with
the previously reported tidal stream of the Sagittarius dwarf galaxy.

The sample is tabulated in electronic form in the online version of
this article, or by request to the authors.

\end{abstract}

\keywords{Galaxy: halo --- Galaxy: structure --- stars: horizontal branch}

%%%% end front matter

\section{Introduction}

Blue horizontal branch (BHB) stars have long been important objects
of study as tracers of the Galactic potential 
\citep{pier_1983,sommerlarsen_christensen_carter_1989,clewley_etal_2002}.  
Their constant absolute magnitude allows accurate
space positions to be determined.  Historically, many researchers
have used samples of BHB stars, or other standard-candle stars such
as RR Lyraes,
to study the local halo, i.e.~the
vicinity of the solar neighborhood where typical distances are of
order the solar Galactocentric radius $R_0 = 8 \kpc$.  These samples
can be used to constrain the kinematics of the local halo
\citep{pier_1984,layden_etal_1996}.
The distant halo (i.e.~typical distances $\gg R_0$), however, is
less well constrained (but see \citet{sommerlarsen_etal_1997}).
Isolating a sample of distant BHB stars would enable study of the structure and
kinematics of the Galactic halo on larger scales
\citep[][hereafter Paper~II]{halo2}, and
estimates of the mass of the Galaxy.

The Sloan Digital Sky Survey (SDSS),
currently halfway through a photometric
and spectroscopic survey of a quarter of the sky, offers an
excellent opportunity to isolate and study BHBs.  
The SDSS spectroscopic survey targets objects between about $g$ = 15.5
to 19 (the bright limit is set to avoid contamination in adjacent
spectroscopic fibers at the slit end of the fiber bundle),
although a significant number of stars outside this range
is also observed. Thus BHB stars
at distances from $\sim 7$ to $\sim 60 \kpc$ from the sun appear in the
spectroscopic survey.  

This paper presents $\nbhb$ BHB stars selected from SDSS that
probe the outer halo. 
The plan of the paper is as follows.  Pertinent characteristics
of the Sloan Digital Sky Survey are summarized in \S\ref{s:sdss}.
\S\ref{s:color_cut} describes the preliminary color cut.  
\S\ref{s:balmer_cut} discusses spectroscopic BHB classification criteria.
For the faint stars, the spectroscopic classification breaks down so
we resort to a more stringent color cut described in \S\ref{s:fainter_stars}.
Radial velocity errors are evaluated in \S\ref{s:duplicates}.
\S\ref{s:absmag} discusses the calculation of bolometric corrections
for accurate distances, and evaluates magnitude errors.
Some sanity checks are presented in \S\ref{s:sanity_checks}, including
an analysis of the limited proper motion data available, and a
map of the Galactic halo which shows the conspicuous
Sagittarius tidal stream (figure~\ref{f:map_3d}).  
The results are summarized in \S\ref{s:summary}.  
The sample is available in Table~\ref{t:data}
in the electronic version of this paper.

\section{The Sloan Digital Sky Survey}\label{s:sdss}
The Sloan Digital Sky Survey \citep[SDSS:][]{york_etal_2000} 
is a project to image
about 1/4 of the sky at high Galactic latitude to a depth of
about 21 to 22 magnitudes in five broad bands
covering the entire optical range and to obtain spectra of about a million
galaxies and 100,000 quasars selected from the imaging data.
While galaxies and quasars are the primary
targets of the spectroscopic survey, the SDSS also obtains a
vast amount of stellar data. Stellar spectra
are acquired by several routes: observations of calibration stars,
observations of various classes of star to be used as ``filler'' observations
when a given region of sky has too low a density of primary targets,
and via objects targeted as candidate galaxies and quasars that turn out to
be stars.

The SDSS uses a dedicated
2.5 meter telescope and a large format CCD camera \citep{gunn_etal_1998}
at the Apache Point Observatory in New Mexico to obtain images
almost simultaneously in five
broad bands ($u,g,r,i,z$) centered at 3551, 4686, 6166, 7480, and 8932 \AA{}
respectively \citep{fukugita_etal_1996,gunn_etal_1998}.
The imaging data are automatically processed through a series of software
pipelines which find and measure objects and provide photometric and
astrometric calibrations to produce a catalogue of objects with
calibrated magnitudes, positions and structure information.  The
photometric pipeline Photo 
\citep[and in preparation]{lupton_etal_2001}
detects the objects, matches the data from the five filters, and
measures instrumental fluxes, positions and shape parameters.  The
last allow the classification of objects as ``point source'' (compatible
with the  point spread function [psf]), or ``extended.''
The instrumental fluxes are calibrated via a network of primary and
secondary stellar flux standards to $\rm AB_{\nu}$ magnitudes 
\citep{fukugita_etal_1996,hogg_etal_2001,smith_etal_2002}.
We use the ``psf magnitude,'' which represents the magnitude
of the best-fit psf to the object, as determined by Photo version
5.4.25 (2003 June). 
The psf magnitudes are currently accurate to about 2\% in 
$g,r$, and $i$ and 3-5\% in $u$ and $z$
for bright ($<$ 20 mag) point sources.  
The SDSS is 50\% complete for point sources to 
$(u,g,r,i,z) = (22.5,23.2,22.6,21.9,20.8)$, and the full-width at
half-maximum (FWHM) of the psf is about $1.5''$ \citep{abazajian_etal_2003}.
The data are saturated at about
14 mag in $g$, $r$ and $i$ and about 12 mag in $u$ and $z$.
Astrometric calibration is carried out using a second set of less
sensitive CCDs in the
camera, which allows the transfer of astrometric catalogue positions
to the fainter objects detected by the main camera.  Absolute positions
are accurate to better than $0.1''$ in each coordinate
\citep{pier_etal_2003}.

The SDSS spectroscopic survey is carried out by targeting objects
selected from the imaging catalogues according to pre-determined and
carefully tested criteria \citep{stoughton_etal_2002}. 
The targeted stars, quasars and galaxies
have limiting magnitudes of around $\rm 18~-~20$ mag
\citep{richards_etal_2002,eisenstein_etal_2001,strauss_etal_2002}.
BHB stars appear in the SDSS spectroscopic sample via two main routes. At
redshifts around 2.7, the quasar color locus crosses the BHB color locus
\citep{richards_etal_2002} and many BHB stars are targeted as candidate
quasars.  BHB stars are also photometrically selected to serve as filler
observations in regions of the sky with a low density of primary targets.
The sky locations of the target objects are
mapped to a series of plug plates which feed a pair of double
spectrographs (A. Uomoto, S. Smee et al.~unpublished) via optical
fibers.  The spectrographs jointly observe 640 objects (including
sky positions and standard stars) simultaneously, and provide
continuous wavelength coverage from about 3800 \AA{} to 9500 \AA{} at
a resolution of about $1800~-~2100$.  The nominal total exposure time for
each spectroscopic plate is 45 minutes (15 minutes $\times$ 3 exposures), 
plus observations of flat
field and arc standard lamps for every plate. The entrance aperture
of the fibers is $3''$.
The spectra are matched to the photometric objects using a video
mapping system for the plug plates (D. Schlegel and D. Finkbeiner,
unpublished).
The spectra are recorded on CCD detectors.  The data are optimally extracted,
the sky spectrum is subtracted, and wavelength, flat-field and
spectrophotometric calibrations are applied automatically by the
Spectro-2D software pipeline, written by D. Schlegel and S. Burles
(unpublished).

Each spectroscopic plate is a circle of diameter $3^{\circ}$ on the sky.
The present work is based on the data of $\nplatesuniq$
unique spectroscopic plates taken to date, through \lastplatedate.  
In addition, $\nplatesredund$ 
plates were observed redundantly, but in defining the BHB sample
we exclude redundant
observations by adopting the plate with the highest S/N.
However, we use the information from these redundant observations
in \S\ref{s:duplicates} below to evaluate velocity errors.

The radial velocity is found by fitting the spectrum to template
spectra, using an automated fitting program
({\bf Spectro\_Brightstars}) written by D. Schlegel
(unpublished).  The basic procedure is as follows.  
\citet{prugniel_soubiran_2001} have published a 
catalogue of about a thousand spectra of stars
with a wide range of properties (spectral types, metallicities etc.)
observed at high spectral resolution with the ELODIE spectrograph.  One
version of the ELODIE spectra 
is binned to R $\sim$ 20,000 and spectrophotometrically
calibrated to about 5\%.  {\bf Brightstars} was used to select template
spectra from the SDSS data which match the ELODIE spectra.  These
template spectra are typically composites from many SDSS spectra, and
are set up to provide relatively coarse sampling of the spectral type
grid.  The template spectra are aligned to zero heliocentric velocity.
Each SDSS spectrum is then compared with the template
spectra in wavelength space,
and the template and wavelength shift that give the best match
is automatically measured.  
The stars in the present sample were
almost all best-fit with spectral types between B6 and A0.
The radial velocities are corrected to the heliocentric standard of rest,
and throughout this paper are denoted $v_r$.

\section{Color cut}\label{s:color_cut}
\begin{figure}
\includegraphics[width=\textwidth]{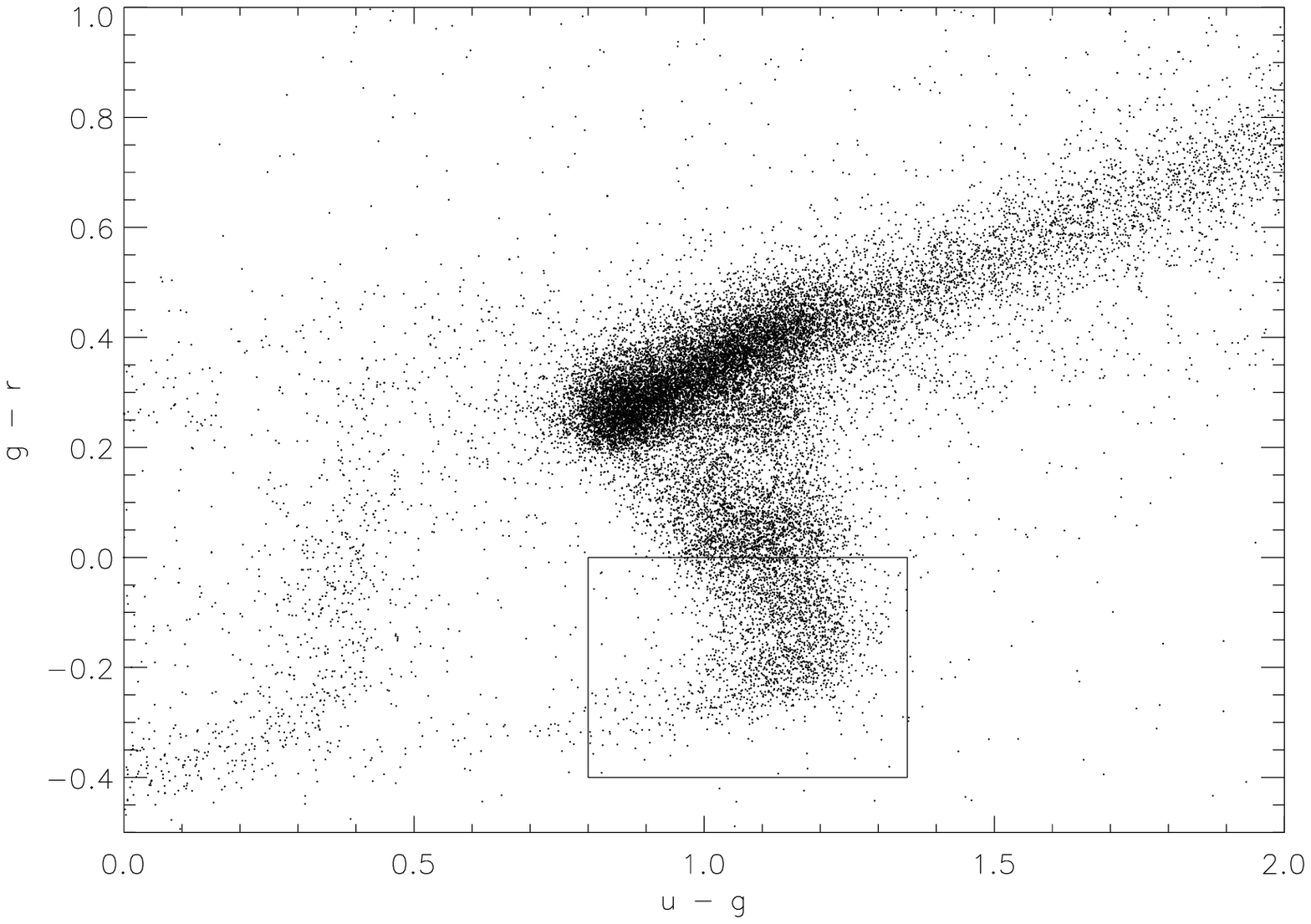}
\caption{SDSS color-color diagram showing all
spectroscopically targetted
objects brighter than $g = 18$ (for clarity)
which were subsequently confirmed as stars. The main sequence
runs from the center of this diagram towards the upper right, and the
large Balmer jump of A-colored stars places them in the offshoot,
where our ``color cut'' selection box is drawn.  
Note that the relative densities of stars in this plot
are affected by selection bias; for example, more stars appear here which
lie in the more ``interesting'' regions of color-color space for QSOs. }
\label{f:ugr_g18}
\end{figure}

BHB stars have very characteristic colors in the 
high-Galactic latitude stellar
sample observed by the SDSS imaging survey, and photometric selection of
candidate objects is straightforward.  BHB stars are bluer in $g-r$
than most halo stars
because the halo main sequence turnoff lies at spectral type F/G.  
The large Balmer jump in
BHB low-gravity stars gives them redder $u-g$ colors than other
blue objects such as low redshift quasars and white dwarfs, and indeed the SDSS
$u$ and $g$ filters are specifically designed to provide optimal
photometric separation of low-redshift quasars
\citep{gunn_etal_1998,york_etal_2000,richards_etal_2002}.  
The SDSS $g-r$ vs.~$u-g$
color-color diagram in figure~\ref{f:ugr_g18} shows all objects in
the spectroscopic sample of $\nplatesuniq$ 
plates used herein 
that were subsequently identified as
stars by {\bf Brightstars}, which is accurate
enough that quasars and galaxies are virtually nonexistent in the
sample.  (Throughout
this paper all magnitudes are corrected for Galactic extinction using the maps
of \citet{schlegel_finkbeiner_davis_1998}.)
The region occupied by BHB stars is indicated with a ``color cut box,''
with boundaries $0.8 < u-g < 1.35, -0.4 < g-r < 0.0$.
The boundaries of this selection box are similar to those used by
\citet{yanny_etal_2000}, and figure~\ref{f:ugr_bhb18} 
(see \S\ref{s:fainter_stars}) shows that our final
sample of bright BHB stars
is well contained within the boundaries.  Stars that lie in this
region of color-color space pass what we will refer to as 
the ``color cut.''

Although BHB stars are effectively selected through the color cut
described above, there remains a significant degree of contamination
from two other types of star.  The locus of cooler F stars is actually
quite outside of the color cut box, to the red in $g-r$
\citep{yanny_etal_2000}, but there are so
many F stars that they leak into the BHB candidate 
sample, via intrinsic variations and photometric errors,
with numbers comparable
to the number of BHB stars.  To identify these cooler stars merely
requires a measure of temperature independent of $u - g$ or
$g - r$.
Somewhat more challenging to identify are the
A-type stars with higher surface gravity
\citep{preston_sneden_2000,yanny_etal_2000,clewley_etal_2002}. These
are probably field blue stragglers, which are some 2 magnitudes
less luminous \citep{preston_sneden_2000}. While the
Balmer jump as measured by $u-g$ can separate out many of these
stars, some remain in the sample.  They can be removed using their broader
Balmer lines 
as described in the next section.

\section{Balmer line analysis cuts}\label{s:balmer_cut}
\begin{figure}
\includegraphics[width=\textwidth]{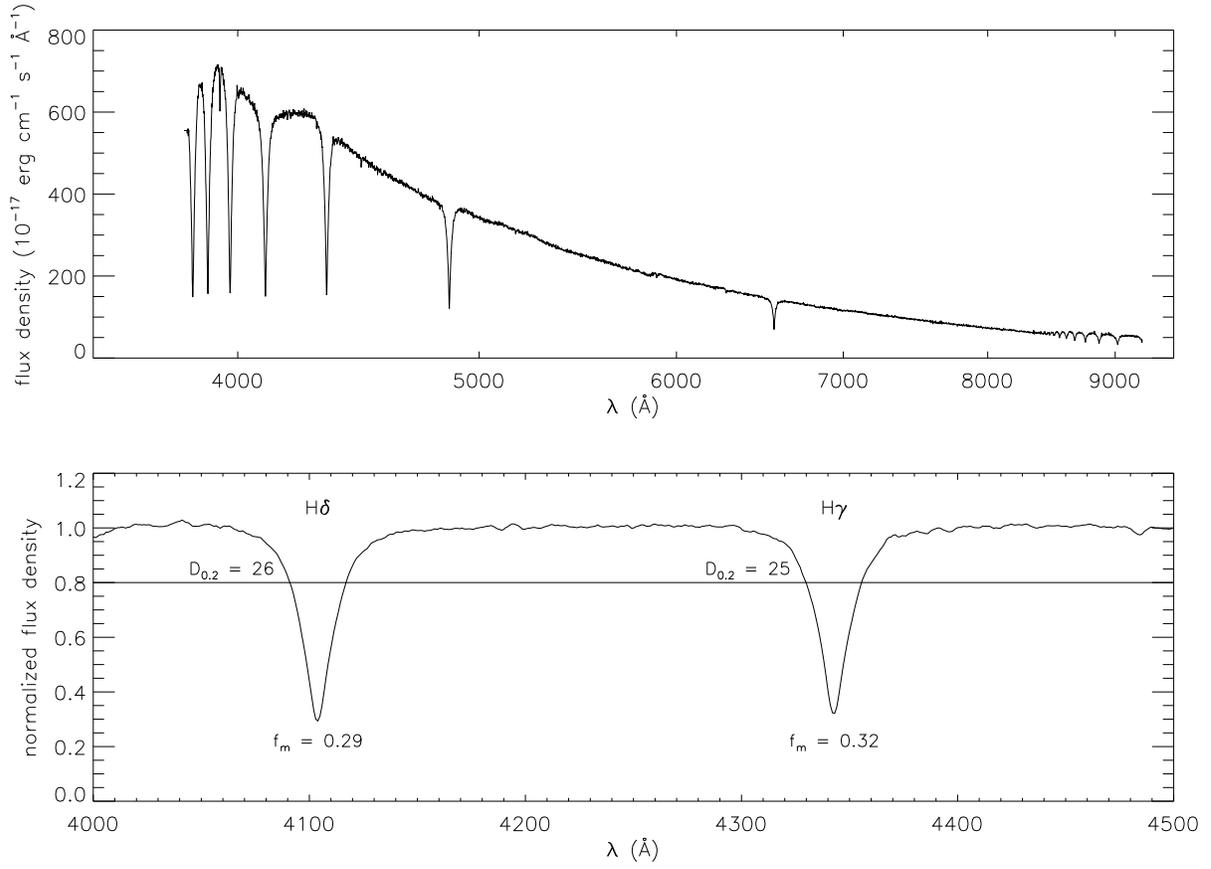}
\caption{Spectrum of a typical (high-S/N) BHB star (upper panel) and the
H$\gamma$-H$\delta$ region of the same star with the continuum
divided out (lower panel).  The parameters $(\fmx,D_{0.2})$ are labelled for
both lines.}
\label{f:bhb_spectrum}
\end{figure}

The upper panel of figure~\ref{f:bhb_spectrum}
presents a typical BHB spectrum, showing
the deep Balmer lines of an A-type star.  In this section we describe
two methods for analyzing these lines: the widely used $D_{0.2}$
method \citep{pier_1983,sommerlarsen_christensen_1986,
arnold_gilmore_1992,flynn_sommerlarsen_christensen_1994,
kinman_suntzeff_kraft_1994,wilhelm_beers_gray_1999} 
and the scalewidth-shape method
\citep{clewley_etal_2002}.  We focus on the H$\gamma$ and
H$\delta$ lines
because the the H$\beta$ and H$\alpha$ lines have lower continuum levels,
and the lines above H$\delta$ are much more closely
spaced, making the determination of the continuum difficult
\citep{yanny_etal_2000}.

To analyze the lines, the continuum must first be fitted and divided out.
For the $D_{0.2}$ method, following \citet{yanny_etal_2000}, we take
the continuum to range from 4000 to $4500 \ang$ with $60 \ang$ masks
covering the positions of the H$\gamma$ and H$\delta$ lines.  We then
divide the spectrum by the best-fit sixth-order Legendre polynomial
to this extracted continuum.  Finally, the spectrum is boxcar-smoothed
with a boxcar length of 5 pixels.
An example normalized and extracted SED is
shown in the lower panel of figure~\ref{f:bhb_spectrum}.
Similarly, for the scalewidth-shape method, we take the continuum
extraction rule of \citet{clewley_etal_2002} (which ranges from 3863
to $4494 \ang$) and fit to a fourth-order Legendre polynomial. After
the continuum is divided out with this prescription,
the SED looks very similar
to the example SED in the lower panel of figure~\ref{f:bhb_spectrum}.

The following terminology is used in this section and throughout.
A ``method'' refers to a general algorithm, which may or may 
not be desirable for isolated use, which employs one or more
``criteria'' for discriminating BHB stars.
The $D_{0.2}$ method and the
scalewidth-shape method are presented in the following.
A ``cut'' is a method or combination
of methods that we \emph{actually} use to eliminate contaminants from
the sample. The combination cut discussed below 
incorporates the $D_{0.2}$ and scalewidth-shape methods.

\subsection{The $D_{0.2}$ method}\label{s:d02_method}
\begin{figure}
\includegraphics[width=\textwidth]{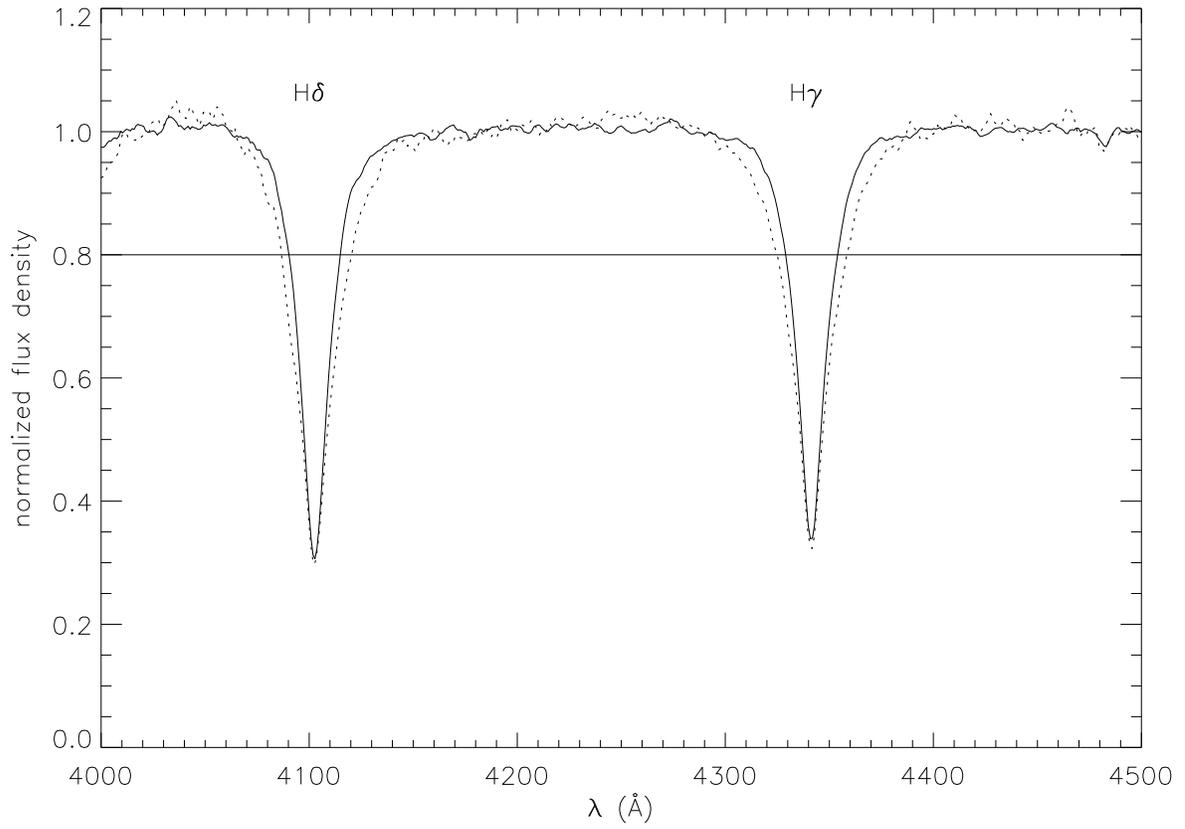}
\caption{The extracted and normalized spectrum of a BHB star (solid)
and a BS star (dotted) clearly showing the BS star's wider Balmer lines
at 80\% of the continuum.}
\label{f:d02_demo}
\end{figure}

Since BHB stars have low surface gravity, their Balmer lines are
narrower than the smaller main sequence blue stragglers.  The
$D_{0.2}$ method discriminates BHB stars from BS stars by determining the
value of $D_{0.2}$, the 
width in angstroms of the Balmer line at 80\% of the continuum
\cite[][]{yanny_etal_2000}.
Figure~\ref{f:d02_demo} demonstrates how this parameter can
discriminate a BHB from a BS star.  Although simple,
this method has been found to be fairly robust and can provide
a better measurement of the linewidth than other more complicated
procedures \citep[e.g.][]{flynn_sommerlarsen_christensen_1994}.

\begin{figure}
\includegraphics[width=\textwidth]{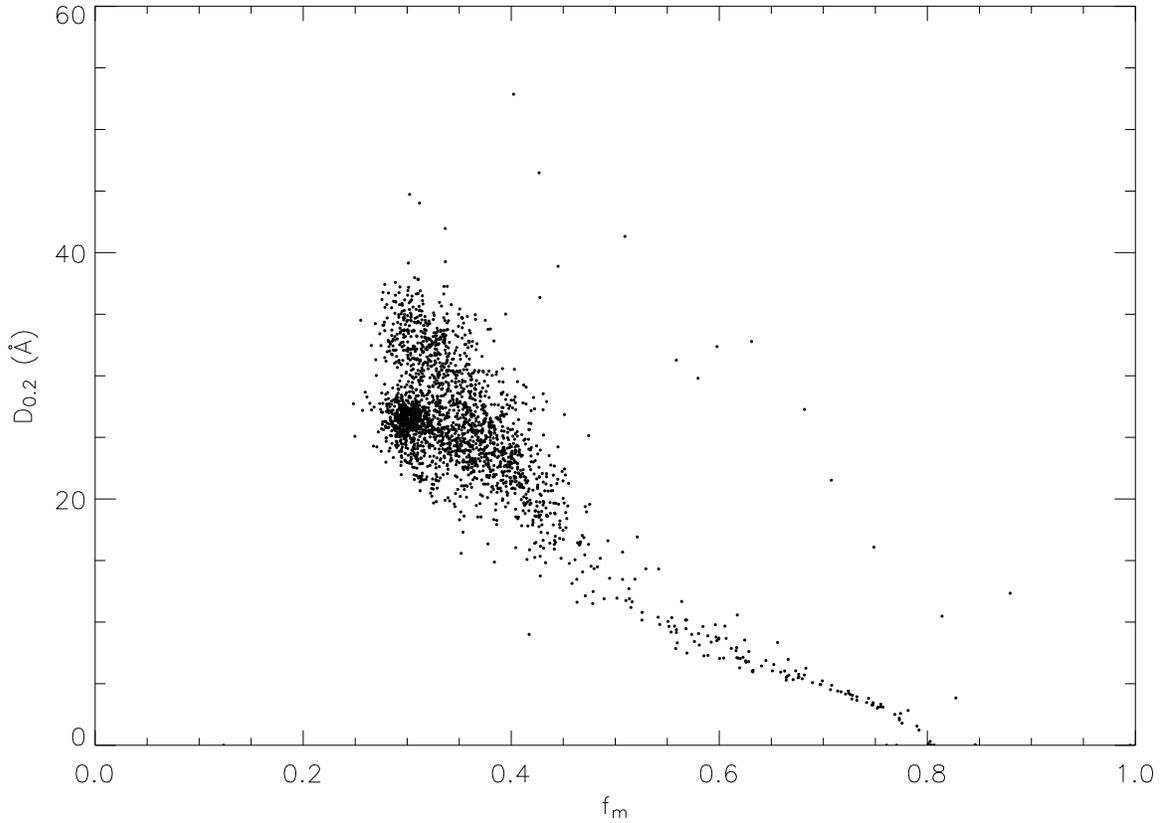}
\caption{The parameters $\fmx$ and $D_{0.2}$ as determined by the
H$\delta$ $D_{0.2}$ method for stars that pass the color cut
with $g < 18$.  The trail of stars with $\fmx \gtrsim .4$ are
too cool to be BHB stars, and the concentration of stars with
$D_{0.2} \gtrsim 28$ is due to blue stragglers with higher surface
gravity. The $\sim 10$ stars that lie well above the main locus for
$\fmx \gtrsim .4$ are placed there by poor parameter determinations
due to missing spectroscopic data at the location of the H$\delta$ line.}
\label{f:d02_v_fm}
\end{figure}

The $D_{0.2}$ method nominally only applies to objects with A-type spectra.
However, as mentioned in $\S$\ref{s:color_cut}, a large contaminant
of our sample is F stars.  Since F-type stars have weaker
Balmer lines than A-type stars, one way of measuring temperature
is by measuring the depth of the H$\gamma$ or H$\delta$ line.
Here we define $\fmx$ as
the flux relative to the continuum at the minimum of the line.  
Thus, higher values of $\fmx$ for the Balmer lines signify
cooler stars.
A plot of $D_{0.2}$ vs.~$\fmx$
for stars passing the color cut is shown in
figure~\ref{f:d02_v_fm}.  In this figure, only stars with $g$ magnitudes
$< 18$ are shown, so that the trends in the plot are not obscured by
the noise in the data of fainter stars (see figure~\ref{f:g18_breakdown},
to be discussed in \S\ref{s:criteria}).  

The concentration of stars near $(\fmx,D_{0.2}) = (.30, 26\ang)$
represents the BHB stars, and those stars with larger values of
$D_{0.2}$ are BS stars.
Figure~\ref{f:d02_v_fm} also suggests that the locii of BS and BHB
stars propagate to larger $\fmx$ and smaller $D_{0.2}$ until the two
locii converge at about $(\fmx,D_{0.2}) = (.4,23\ang)$.  In other words,
for a given $.3 \lesssim \fmx \lesssim .4$,
the distribution of stars in $D_{0.2}$
is bimodal, but the stars are most cleanly separated for the smaller
values of $\fmx$ and not separated at all when $\fmx \sim .4$.
Even though figure~\ref{f:d02_v_fm} doesn't show the trend clearly, we return
to this point in \S\ref{s:criteria} and show that the stars
with smaller values of $D_{0.2}$ but with $.34 \lesssim \fmx \lesssim .4$ are
the other horizontal branch stars; RR Lyraes, for example, 
fall into this region of the horizontal branch.
Because horizontal branch stars are evidently more difficult to 
discriminate from main sequence stars
at lower temperatures, we adopt the criteria $\fmx < .34, D_{0.2} < 28$
for likely BHB stars for the H$\delta$ $D_{0.2}$ method.

\subsection{The scalewidth-shape method}

\citet{clewley_etal_2002} have proposed the scalewidth-shape method
for discriminating BHB stars based on the S\'ersic profile for Balmer
lines \citep{sersic_1968}:
\begin{equation}
y = 1.0 - a \exp{\left[-\left(\frac{|x-x_0|}{b}\right)^c\right]}.
\end{equation}
Here, $y$ is the normalized flux and $x$ is the wavelength in $\ang$.
We fit the normalized extracted spectrum (where the bounds of extraction
are given by \citet{clewley_etal_2002}) to the S\'ersic profile with
three free parameters: $a,b$ and $c$.  The parameter $x_0$ is assumed
to be the nominal location of the Balmer line corrected for redshift, which
has already been determined.  After fitting both Balmer lines of
every star to the
three-parameter ($a,b,c$) S\'ersic profile, there is a reasonably small
dispersion in the determination of $a$: for the $H\gamma$ line the
average is $.690$ and the root-mean-square (rms) is $.019$; for 
$H\delta$ the average is $.740$ and the rms is $.018$.
We fix $a$ at its average value among stars passing the BHB criteria of the
H$\delta$ $D_{0.2}$ method, which is not objectionable because the
$D_{0.2}$ method is independent of the scalewidth-shape method.
The Balmer lines are then refitted to the
two-parameter ($b,c$) S\'ersic profile.
This refitting is important because errors
in $a$ are correlated with the errors in $b$ and $c$
\citep{clewley_etal_2002}.  For this work we adopt $a_\gamma = .690$
for the H$\gamma$ line and $a_\delta = .740$ for the H$\delta$ line.
The optimum value of $a$ most likely depends on the resolution of
spectroscopy and similar factors.

Fitting to the S\'ersic profile thus provides the two parameters $b$ and $c$
for both the H$\gamma$ and H$\delta$ lines, for every star.
\citet{clewley_etal_2002} show that BHB and BS stars separate rather
cleanly in a plot of $b$ against $c$.  Furthermore, they show that
$c$ is a measure of temperature: cooler stars have smaller values of $c$.

\subsection{Determination of criteria for the $D_{0.2}$ 
	and scalewidth-shape methods}\label{s:criteria}

We concentrate on the set of brighter stars with $g < 18$ because
the noise in the spectra of fainter stars throws large errors in
the parameter determinations for the $D_{0.2}$ method and the
scalewidth-shape method.  This point is illustrated in
figure~\ref{f:g18_breakdown}.  We will return to the selection of
BHBs from the fainter star sample in \S\ref{s:fainter_stars}.

\begin{figure}
\includegraphics[width=\textwidth]{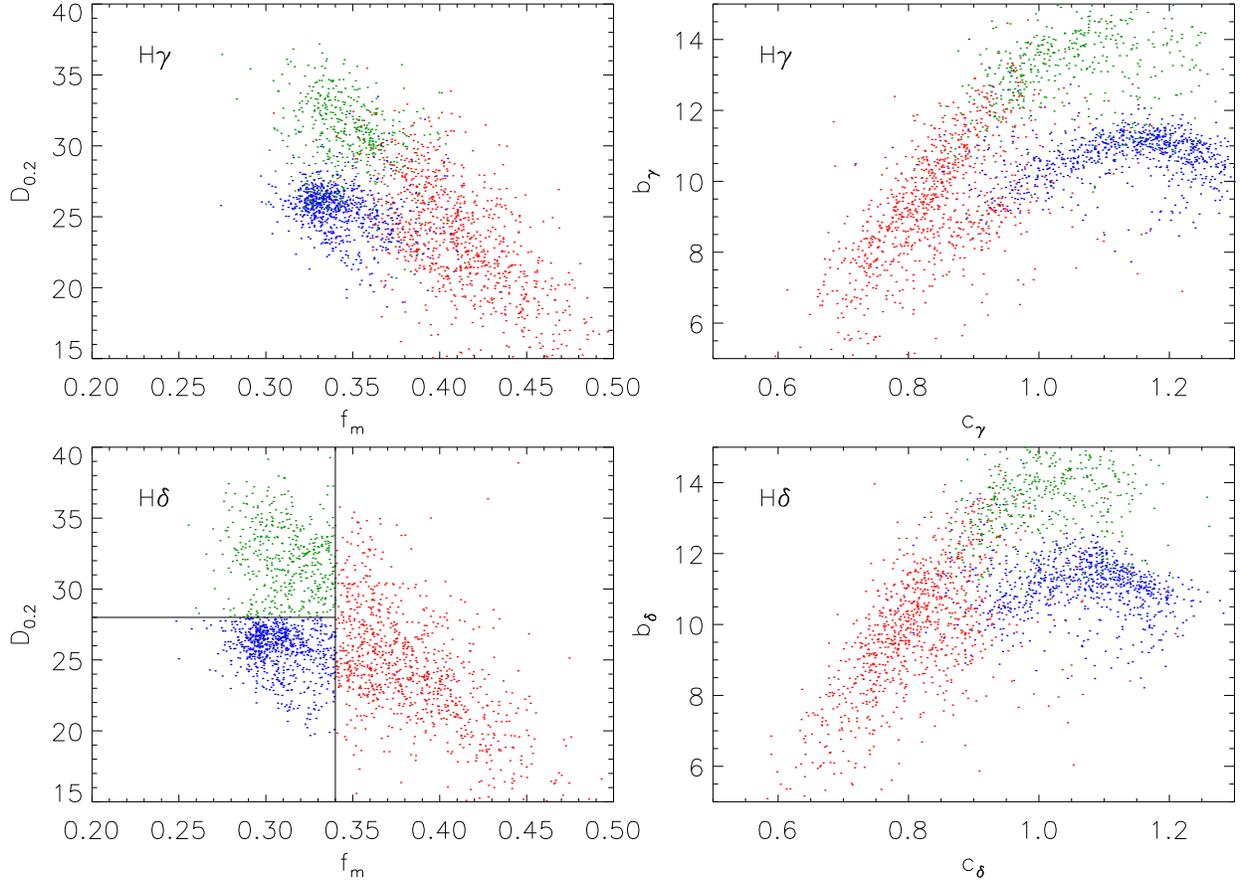}
\caption{Comparison between the $D_{0.2}$ method (left panels) and
the scalewidth-shape method (right panels) for the H$\gamma$ line
(top panels) and the H$\delta$ line (bottom panels).
Each star is represented by the same color in all four plots, so
the compatibility among all four methods is seen.  Here stars are colored
according to their classification by the $D_{0.2}$ method for
the H$\delta$ line: BHB stars are blue, BS stars are green, and
the cooler stars are red. Only stars with $g < 18$ are plotted
here.}
\label{f:compare1}
\end{figure}

Figure~\ref{f:compare1} shows the parameters of the sample of BHB
candidates with $g < 18$
as determined by the two linewidth analysis methods described in the
preceding sections.  Each star
is represented by the same color in all four panels.
In this figure the H$\delta$ $D_{0.2}$ method is represented in
the lower left panel, and stars
which pass the BHB criteria for this method are colored blue,
stars which are likely blue stragglers
are green, and all others (cooler stars) are red.  It is readily
apparent that the four \emph{independent} parameter spaces map BHB
stars, BS stars, and cool stars to different regions.  

\begin{figure}
\includegraphics[width=\textwidth]{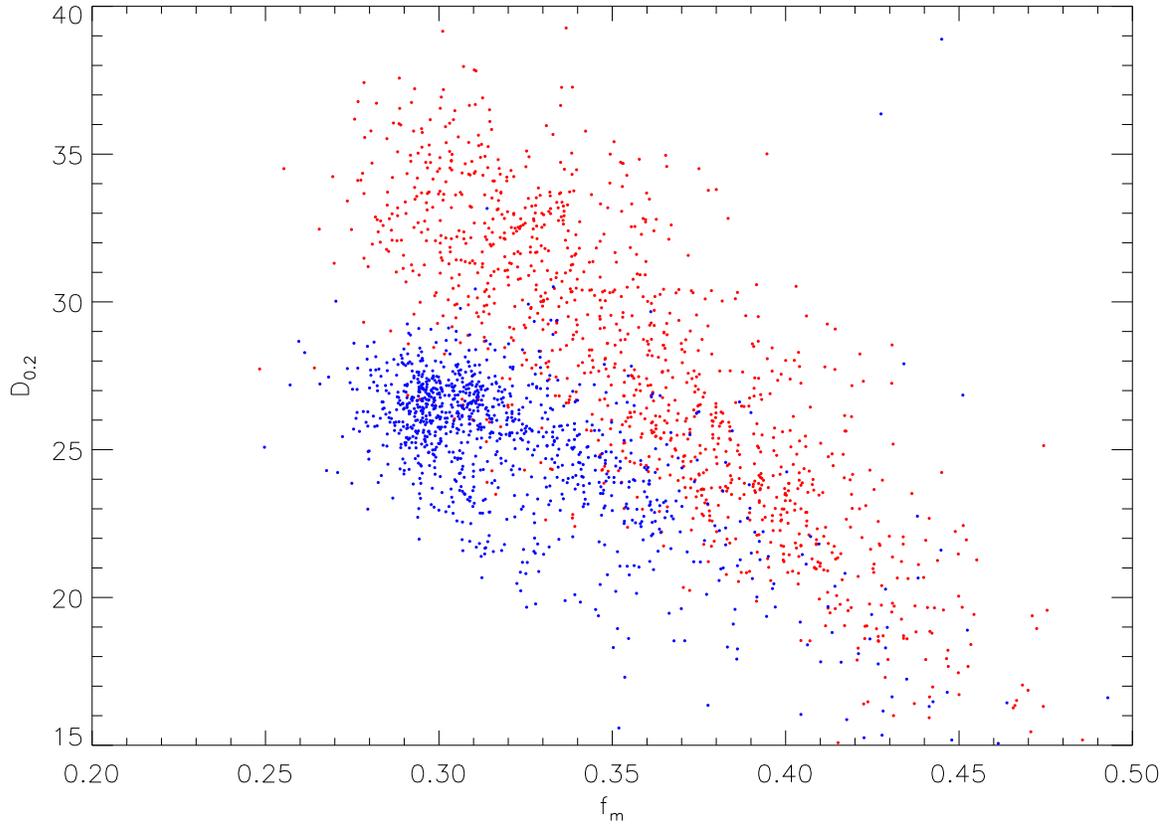}
\caption{Parameter determinations from the H$\delta$ $D_{0.2}$ method
for stars with $g < 18$.  Stars are colored according to their
location on the $b-c$ diagram for the H$\gamma$ scalewidth-shape method:
stars for which $b < 12 - 30(c-1.18)^2$ are colored blue; all others are
colored red.  (The temperature criterion is not used.) Thus stars which
have low values of $b$ from the scalewidth-shape method also have
low values of $D_{0.2}$, and are likely horizontal branch stars. }
\label{f:compare_rhb}
\end{figure}

The exact BHB criteria for each method were determined through
a combination of adjusting
the bounds of the criteria by eye and comparing the results of one method to
the others.  For instance, the scalewidth-shape method
separates BHB and BS stars cleanly enough that there is a natural
gap between the two distributions in a plot of $b$ vs.~$c$, as shown in
figure~\ref{f:compare1}.  A parabola is qualitatively
drawn to divide the two distributions.  Similarly, with regard to
the $D_{0.2}$ method, the distribution in $D_{0.2}$ of the stars
with low values of $\fmx$ naturally separates into the two populations
of BHB and BS stars.  The issue of where to draw the temperature criterion
($\fmx$ for the $D_{0.2}$ method and $c$ for the scalewidth-shape
method) is more sensitive because the ``natural'' gap is harder to see
in the $D_{0.2}$ method and seems to be nonexistent in the
scalewidth-shape method.  In fact, the
upper right panel of figure~\ref{f:compare1} hints that the division
between stars with low surface gravity and high surface gravity
continues even for much cooler stars than BHB stars.  The
cooler population ($0.8 \lesssim c_\gamma \lesssim 1.0$) with lower
surface gravity (lower $b_\gamma$) is thus the horizontal
branch stars which are not blue, e.g.~RR Lyraes.  
This $b$ vs.~$c$ plot corroborates
the evidence in \S\ref{s:d02_method} that non-blue
horizontal branch stars form a locus in the $(\fmx,D_{0.2})$ plane which
extends out to cooler temperatures corresponding to $\fmx \sim .4$.
Figure~\ref{f:compare_rhb} shows that the stars of the population
that has smaller values of $b_\gamma$ also have smaller values of
$D_{0.2}$.
With this information, we deduce that the BHB stars are located
in the clump in the H$\delta$ $\fmx-D_{0.2}$ plane centered on
$(\fmx,D_{0.2})_{\rm{H}\delta} \sim (.30,26\ang)$ and we set the temperature
criterion at
$\fmx = .34$.  A subsequent inspection of figure~\ref{f:compare1}
reveals that most BHB stars have $c_\gamma > 1.0$, so we adopt this
as the temperature criterion for the H$\gamma$ scalewidth-shape method.
Table~\ref{t:criteria} lists the adopted BHB selection criteria for
each method.

\subsection{The combination cut}
	\label{s:combination_cut}

We define the combination cut as those stars which pass the criteria for 
(the H$\gamma$ $D_{0.2}$ and H$\delta$ scalewidth-shape methods) 
OR (the H$\delta$ $D_{0.2}$ and H$\gamma$ scalewidth-shape methods).  
We assume that the H$\gamma$ $D_{0.2}$ and H$\delta$
scalewidth-shape methods are completely independent of each other, as 
are the H$\delta$ $D_{0.2}$ and H$\gamma$ scalewidth-shape methods.
This assumption is reasonable 
because no information is shared in the analyses of either line, and
the $D_{0.2}$ and scalewidth-shape methods measure fundamentally different
properties of the spectral energy distribution.
Under this assumption, the contamination fraction 
for the combination cut would be 
$\kappa_{\rm{combo}} \approx 2\kappa^2$, where $\kappa$ is the
contamination fraction of a single method.
In Appendix~\ref{s:contamination} we show that the contamination
fraction of any one of the four methods is probably $\kappa \lesssim 10\%$,
so $\kappa_{\rm combo} \lesssim 2\%$ under these assumptions.
However, to be more conservative, the fact that 
$\kappa \sim 10\%$ for all four individual methods suggests that
$\kappa_{\rm{combo}}$ is probably $10\%$ or less; in Paper~II
we use $\kappa_{\rm combo} = 10\%$ to study errors introduced 
by contamination into the kinematic analysis .

\section{The fainter stars}\label{s:fainter_stars}
\begin{figure}
\includegraphics[width=\textwidth]{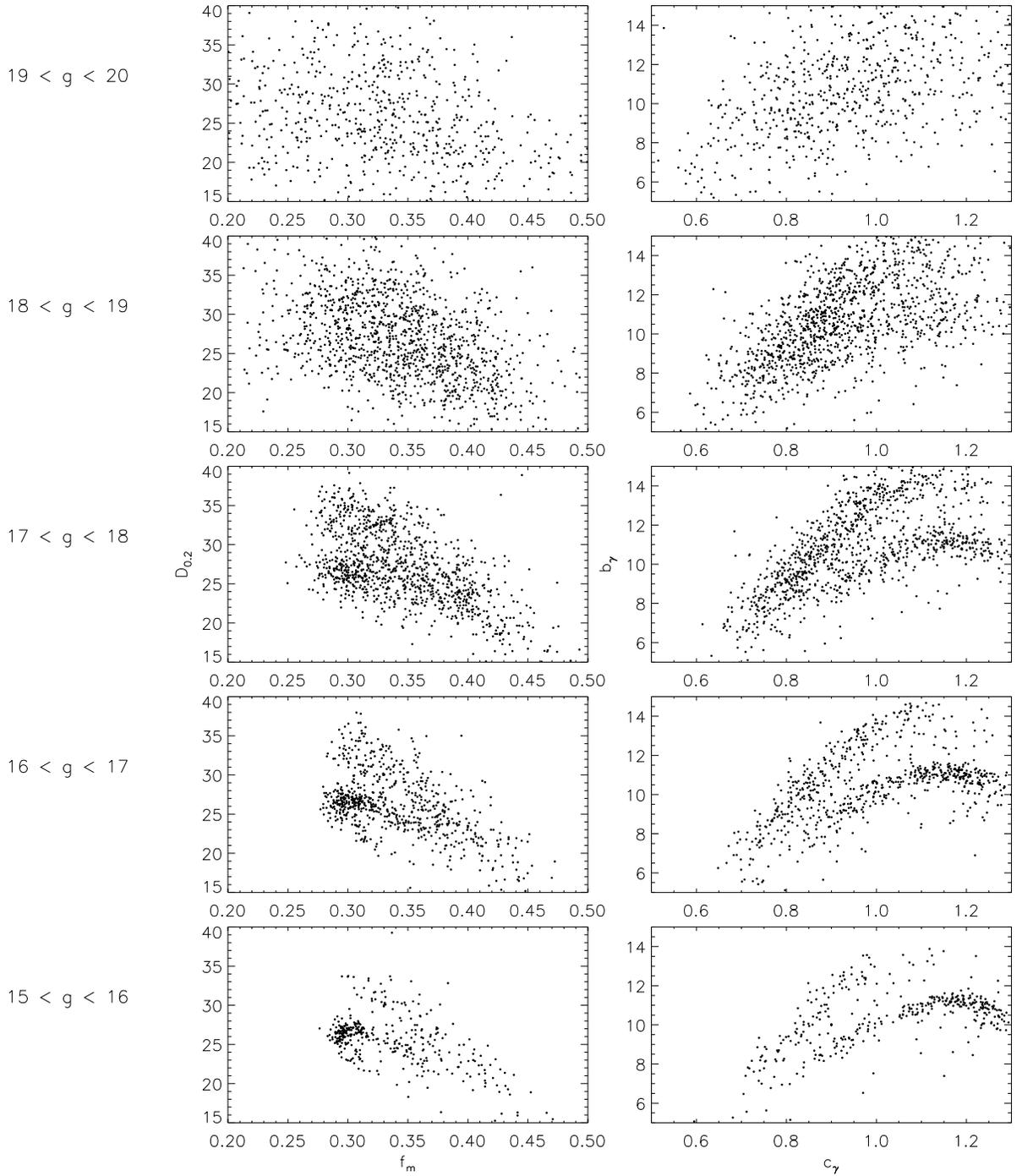}
\caption{Parameter determinations of the H$\delta$ $D_{0.2}$ method 
(left panels) and of the H$\gamma$ scalewidth-shape method (right
panels) for five $g$ magnitude bins, labelled.  Errors evidently 
become too large to safely discriminate BHB stars from other types 
of stars at about $g \sim 18$.}
\label{f:g18_breakdown}
\end{figure}

\begin{figure}
\includegraphics[width=\textwidth]{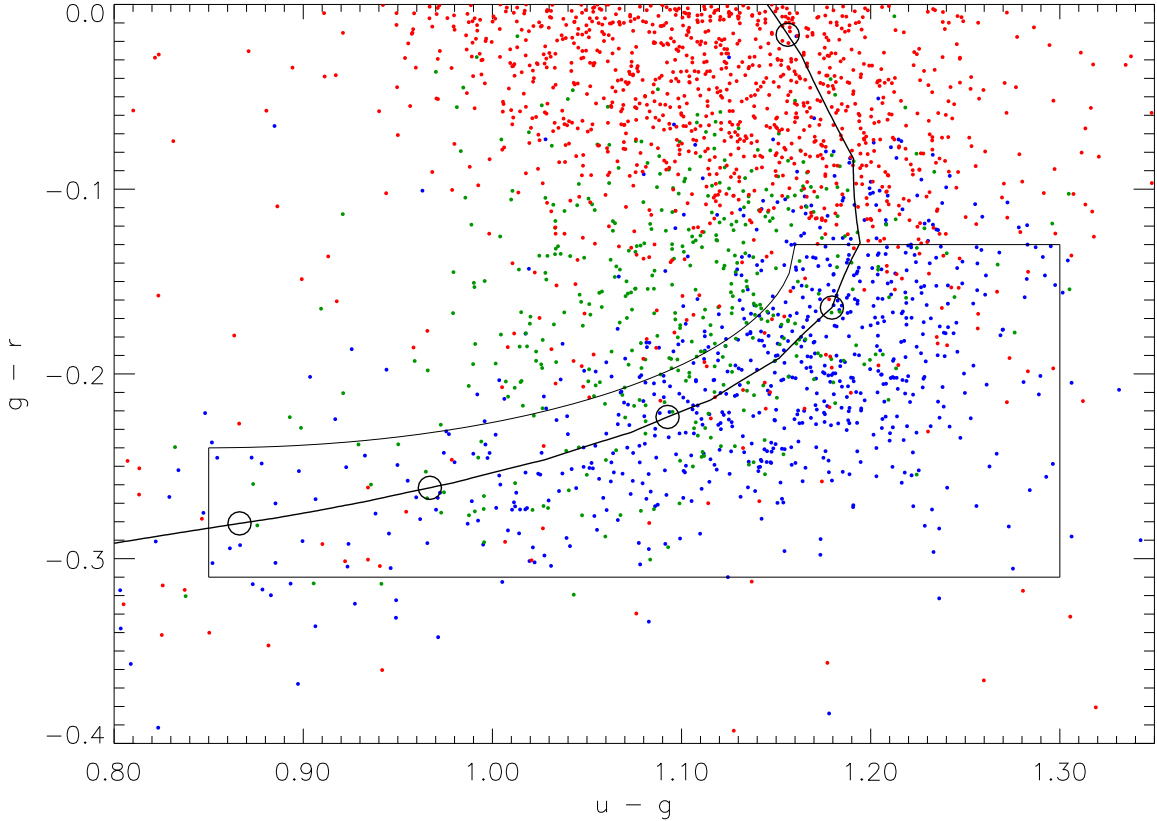}
\caption{The same color-color diagram presented in figure~\ref{f:ugr_g18}
but zoomed into the color cut box and with stars colored
according to their determined classifications by the combination cut.
Only stars with $g < 18$ are shown.
BHB stars are blue, BS stars are green, and all others
(F stars, cooler horizontal branch stars) are red.  Note
that the three samples do comprise different regions of color space,
suggesting that purely photometric separation criteria are possible,
but that the overlap between the regions is quite significant.
The sequence of theoretical horizontal branch models discussed
in \S\ref{s:absmag} is plotted as the thick curved line.  The five
black circles on the thick line denote, starting at the top and progressing
down and to the left, stars with absolute magnitudes 
$g = (0.6,0.55,0.6,0.7,0.8)$.}
\label{f:ugr_bhb18}
\end{figure}

As shown in figure~\ref{f:g18_breakdown}, the linewidth analysis methods
cannot be expected to perform accurately for fainter stars than 
$g \gtrsim 18$; in fact the number of $g > 18$ stars that
are identified differently in any two methods ($\sim 100 - 400$) 
is of the same order as the number of $g > 18$ stars 
identified as BHB stars in one method ($\sim 200 - 500$).  
Because these fainter stars
have noisier spectra, not only are the algorithmic BHB criteria unreliable,
but eye-inspection would be difficult at best.
However, an inspection of the results of the combination cut applied
to the bright ($g < 18$) stars, as shown in figure~\ref{f:ugr_bhb18},
reveals that BHB stars, BS stars, and cooler stars occupy different regions
of color-color space, although there is significant overlap.  In this
figure BHB stars are defined as those which pass the combination cut;
(green) BS stars are defined as those which do not pass the combination cut but
lie in the BS region for the H$\delta$ $D_{0.2}$ method; and red stars
are everything else.  The thick curved line with periodic circles will
be explained in \S\ref{s:absmag}.

The bounds of this particular plot are the same as the color cut discussed in
\S\ref{s:color_cut}, so it is evident that the BHB stars are more
likely to be found in the region enclosed by the ``piano'' shape indicated than
outside of it.  
Since SDSS photometry can be expected to be
accurate to a fainter magnitude \citep[$g \sim 20$,][]{york_etal_2000}
than can the
combination method (see figure~\ref{f:g18_breakdown}), we select
BHB stars with $g > 18$ solely on the basis of color.
The new color cut is indicated by the piano
shape in figure~\ref{f:ugr_bhb18} (and is given by the region
$0.85 < u-g < 1.3, -0.31 < g-r < -.13$ 
excluding the elliptical subregion 
$((u-g-0.85)/0.31)^2 + ((g-r+0.13)/0.11)^2 < 1$) and will
be referred to here as the ``stringent'' color cut.

% to do this analysis use stars_lw_sersic_mouse and assign_cut
What are the contamination and incompleteness for the
stringent color cut with no spectroscopic analysis?  Assuming
that the combination cut is 100\%
accurate for $g < 18$, there are $\nbhbbr$ BHB stars, 
$\nbhbbrnotstringent$ of which
do not pass the stringent color cut.  There are 
$\nnotbryesstringent$ stars that
pass the stringent color cut that are not classified as BHB
stars by the combination method.  Most of these ($\nbsbryesstringent$)
are classified as blue stragglers by the H$\delta$ $D_{0.2}$ method.
Thus the contamination
for the stringent color cut alone is 
$199/(733-172+199) \approx 25\%$.

\section{Duplicate stars and radial velocity errors}\label{s:duplicates}
\begin{figure}
\includegraphics[width=\textwidth]{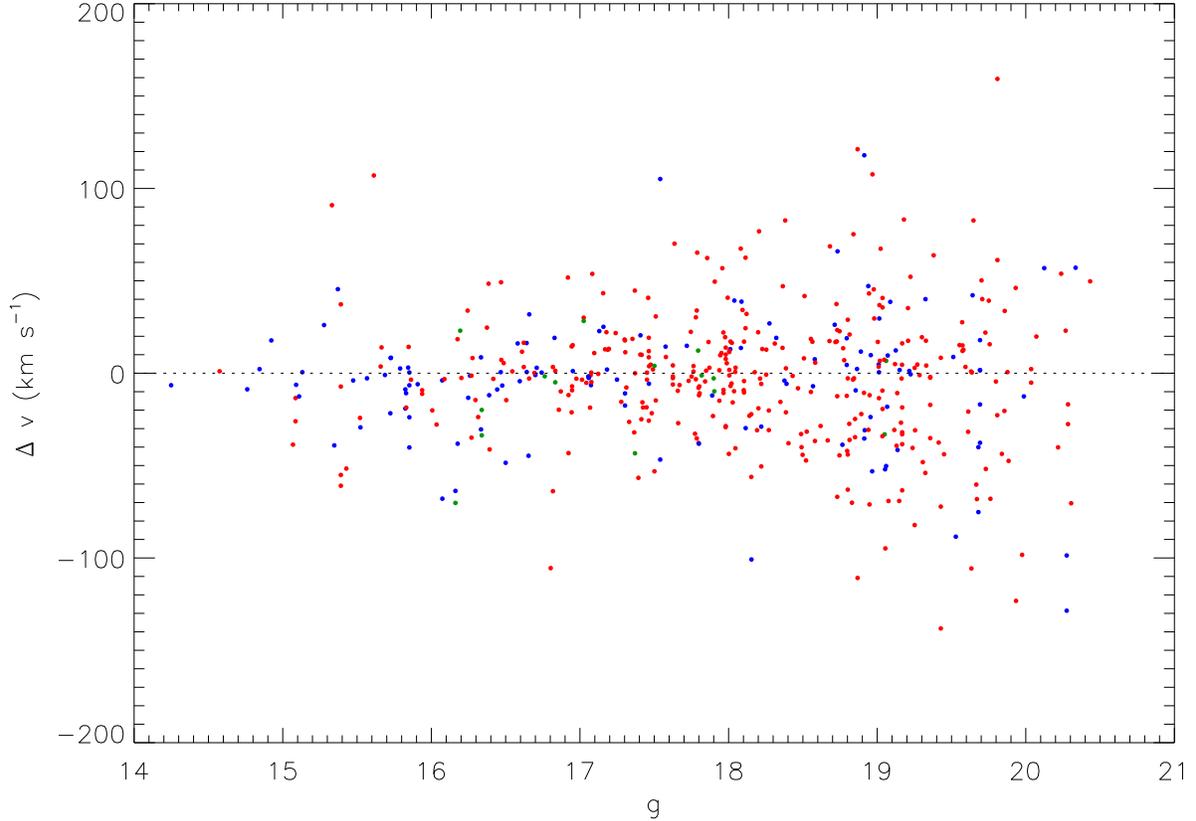}
\caption{Discrepancies between the radial velocity determinations 
$\Delta v_r$ from
duplicate observations of the same stars vs.~$g$ magnitude.
The velocity error increases for faint stars, as expected.  
Stars are colored blue if both observations indicate a BHB star 
(using the combination cut for $g<18$ and the stringent color cut
for $g>18$), red if both indicate a non-BHB star,
and green if the two observations are in conflict.  The reason why there 
are some conflicting classifications for $g>18$ is because
the two spectroscopic observations sometimes actually reference different
photometric observations when available.}
\label{f:duplicates}
\end{figure}

Since the spectroscopic plates overlap on the sky, sometimes objects
are targeted for spectroscopy on two different plates.
Of the $\ncolorcut$ stars passing the color cut (from \emph{unique} plates), 
$\ncolorcutdup$ are duplicate observations of the same star on different plates.
We remove these $\ncolorcutdup$ redundant observations from our BHB sample.
In addition, from the data of the $\nplatesredund$
non-unique plates (see \S\ref{s:sdss}), 
there are $\ndupfromredund$ stars 
passing the color cut which duplicate observations
of stars from the unique plates.  Each spectroscopic observation
of a star provides an independent measurement of radial velocity $v_r$
and, for the bright ($g < 18$) stars,
an independent classification as a BHB star or non-BHB star.
Figure~\ref{f:duplicates} shows the $\nduptotal$ stars which have 
multiple spectra and therefore independent determinations of $v_r$.
The ordinate axis shows the difference in the radial velocity determinations
$\Delta v_r$ between each duplicate star and its corresponding 
star in our purified sample of unique stars.
Stars which are classified as BHBs for both observations are colored blue,
stars which are classified as non-BHBs for both observations are colored red,
and stars which have conflicting classifications between the two observations
are colored green.  Brighter than $g = 18$,
there are $\nbrbhbbhb$ stars which are doubly classified as BHBs, 
$\nbrnotnot$ stars
which are doubly classified as non-BHBs, and $\nbrbhbnot$ stars which have
conflicting classifications.  This is consistent with the $10\%$
contamination determined in Appendix~\ref{s:contamination}, assuming
that there is about an equal number of contaminants in the BHB sample 
as the number of BHB stars which are not classified 
as such (i.e., assuming $\kappa \approx \eta$).
Figure~\ref{f:duplicates} also shows the dependence of velocity errors
on magnitude; as expected, fainter stars tend to have greater velocity errors,
presumably due to noiser spectra.
The rank correlation coefficient of $(g, |\Delta v_r|)$ is $\rankcorr$.
The velocity differences determined this way turn out to be consistent
with the errors derived from the spectral fits for only about half
of the stars, independent of the magnitude or whether or not a star is
a BHB or BS star.  This comparison implies that the errors have been 
underestimated and this is currently under investigation.
The rms of $\Delta v_r$ in 
figure~\ref{f:duplicates} is $\rmsverror \kms$.
Clearly the errors in radial velocity
measurements are not gaussian, but a meaningful approximate result may still
be obtained if they are assumed to be so.  Since the distribution
of $\Delta v_r$ is based on the errors of two measurements, the 
error of one measurement is approximately 
$\Delta v_r \approx \rmsverror/\sqrt{2} \approx 26 \kms$.

\section{The absolute magnitude of BHB stars}\label{s:absmag}

In order to refine our photometric distance estimates,
we have computed theoretical colors and bolometric corrections
for BHBs.  We integrate the latest Kurucz model atmospheres (downloaded
from {\tt www.kurucz.harvard.edu}) against the
SDSS filter response functions at a standard
airmass of 1.2 atmospheres, in the manner described by
\cite{Lenz_etal98}.  Unfortunately, the latter authors report
only colors, so it was necessary to repeat their calculations in
order to obtain the bolometric corrections.
Apart from wavelength, the Kurucz spectra
are functions of three parameters: effective temperature $\Teff$,
surface gravity $g=GM/r^2$, and metallicity relative to solar $[M/H]$.

To derive absolute magnitudes and bolometric corrections, it is
necessary to assume a relationship between the luminosity $L$
and $\Teff$.  The models of \cite{Dorman_etal93} for
the theoretical zero-age horizontal branch (ZAHB) at
$[M/H]\le-1.5$ can be summarized roughly by
\begin{equation}\label{e:ZAHB}
\log(L/L_\odot)\approx 1.58~-0.73\log(\Teff/10^4\K)
\end{equation}
over the range $8000\K\le\Teff\le12000\K$.
From a detailed study of the horizontal branch of M5, 
however, \cite{Baev_etal01} find that luminosity
varies more weakly with $\Teff$; in fact
their results appear to be consistent with constant
luminosity and mass,
\begin{equation}\label{e:MLrels}
M\approx 0.6\,M_\odot,\qquad \log \frac{L}{L_\odot}\approx 1.65
\end{equation}
with a scatter of no more than $\pm0.05 M_\odot$
and $\pm0.05\,\mbox{dex}$, respectively,
over the same temperature range.
We adopt the mean values in equation (\ref{e:MLrels}), so that
$\log g\approx 3.52+4\log(\Teff/10^4\K)$.  

To give an idea of the theoretical uncertainties, we note that if in
fact the relation (\ref{e:ZAHB}) holds for our HB stars, then by
adopting (\ref{e:MLrels}) we have overestimated their distance moduli
by an average of $0.18\,\rm{mag}$ over the indicated range of $\Teff$,
and by neglecting the variation of $L$ with $\Teff$,
we have increased the scatter in distance modulus by $\approx
0.09\,\rm{mag}$, which should be taken in quadrature with the
photometric measurement error.

The results of this exercise are shown in Table~\ref{t:thb}.
Evidently, for the colors of interest to us,
$\Teff\in(8000,12000)\K$, and
the influence of metallicity is slight if $[M/H]\le -1$.

The sequence of theoretical horizontal branch stars, parametrized by
temperature, is projected onto the $(u-g,g-r)$ plane
in figure~\ref{f:ugr_bhb18}.  The five black circles
on the thick line denote, starting at the top and progressing down and 
to the left, stars with absolute magnitudes $g = (0.6,0.55,0.6,0.7,0.8)$.
For every observed BHB star in our sample (colored blue in the figure)
we define the most probable absolute magnitude associated with its
$(u-g,g-r)$ colors by simply finding the absolute magnitude of the
point on the theoretical track that is closest to the observed star in
this color-color space.

The stringent color cut 
(piano shape in figure~\ref{f:ugr_bhb18}, see \S\ref{s:fainter_stars})
was defined before the theoretical track was calculated, so the fact that
the theoretical track satisfies the stringent color cut is a meaningful
consistency check.

\begin{figure}
\includegraphics[width=\textwidth]{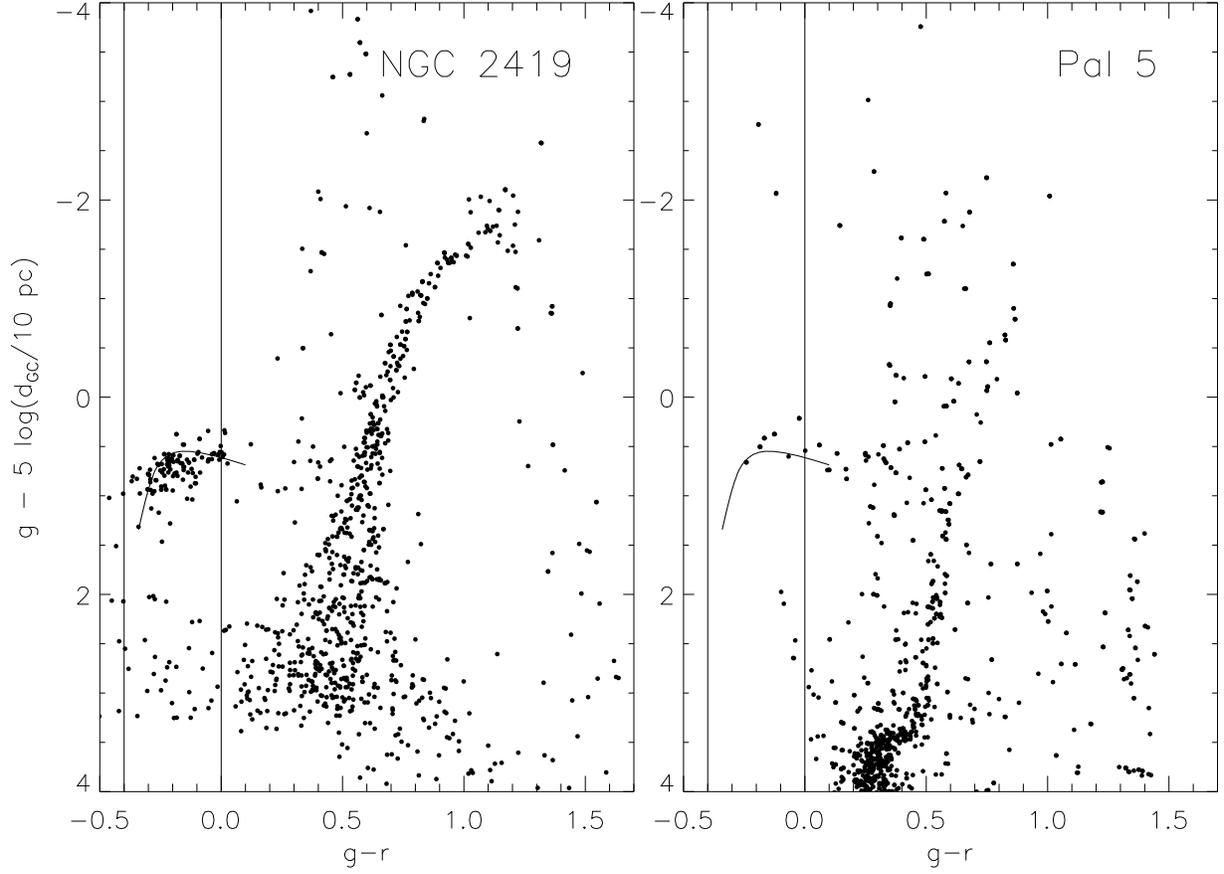}
\caption{Color-magnitude diagrams for the Galactic globular clusters
NGC 2419 and Pal 5.  The distance modulus has been subtracted from
the dereddened $g$ magnitudes to arrive at directly observed
absolute magnitudes.  Lines at $g-r = (-0.4,0.0)$ denote
the color cut.  The theoretical horizontal branch sequence is overplotted.}
\label{f:gc_cmd}
\end{figure}

As a further consistency check, 
we have isolated several Galactic globular clusters
observed by the SDSS photometric survey to date and plotted their
color-magnitude diagrams (CMDs).  
Pal 3, NGC 5904, and NGC 6205 do not show a discernable horizontal branch,
so they are not considered here.  The $g$ vs.~$g-r$ CMDs of Pal 5 
\citep{yanny_etal_2000,ivezic_etal_2000} and NGC 2419 are presented
in figure~\ref{f:gc_cmd}, where the distance modulus, derived from
cluster distances of $d = 84.2$ and $23.2 \kpc$ for NGC 2419 and Pal 5,
respectively \citep{harris_1996}, has been subtracted
from the dereddened apparent magnitude $g$.
Stars within $8'$ of the center of NGC 2419 and within $14'$ of the center
of Pal 5 were selected.  
The SDSS automated photometry avoids analyzing
very crowded fields, so the stars presented here, though within
the stated radii, tend to lie in
the outskirts of the clusters.
Vertical lines at $g-r = (-0.4, 0.0)$ indicate our color cut.
The theoretical horizontal branch sequence is also plotted in both panels.
The horizontal branch stars of NGC 2419 appear to be systematically
fainter than the theoretical prediction, while those of Pal 5 seem
to be systematically brighter; however, these systematic differences
could easily be due to inaccurate distances $d$.
By selecting from these two globular clusters the stars passing the color cut 
($0.8 < u-g < 1.35, -0.4 < g-r < 0.0$) and lying in the region
of the CMD for the horizontal branch 
($0.0 < g - 5 \log{(d/10 \pc)} < 1.5$), we compare
the directly observed absolute magnitude $g - 5 \log{(d/10 \pc)}$
to the indirectly obtained theoretical absolute magnitude based on the 
$(u-g,g-r)$ colors.  For NGC 2419, the rms 
of the differences
of the two among the 94 stars is .134 magnitudes, and the 
root-sum-square (rss) is .167 magnitudes\footnote{
	${\rm rms} = \sqrt{\frac{1}{N} \Sigma \left( 
	\Delta g_i - \langle\Delta g\rangle \right)^2}$,
	$\langle\Delta g\rangle = \frac{\Sigma \Delta g_i}{N}$,
	${\rm rss} = \sqrt{\frac{1}{N}\Sigma \Delta g_i^2}$}.
For Pal 5, the rms is .126 magnitudes and the rss
is .202 magnitudes.  The difference between the rms and rss is due
to the nonzero mean, or systematic error as discussed above.  Therefore,
the magnitude error of a star, with absolute magnitude
determined by comparison to the theoretical horizontal branch sequence
described above, is of order .2 magnitudes.  The error may be as small 
as $\sim .13$ magnitudes if the only cause of the systematic difference
between directly observed absolute magnitudes and our indirectly obtained
absolute magnitudes is incorrect distances for Pal 5 and NGC 2419.

\section{Sanity checks}\label{s:sanity_checks}
As figure~\ref{f:ugr_bhb18} shows, BHB stars determined by the combination
cut are in general redder in $u - g$ and bluer in $g - r$ than
other objects such as blue stragglers.  This result is consistent with
the work of \citet{yanny_etal_2000},
so it is reassuring.  
In this section we present several more sanity checks which verify that
our sample of BHB stars is probably accurate.

\subsection{Metallicity from the CaII K line}
\begin{figure}
\includegraphics[width=\textwidth]{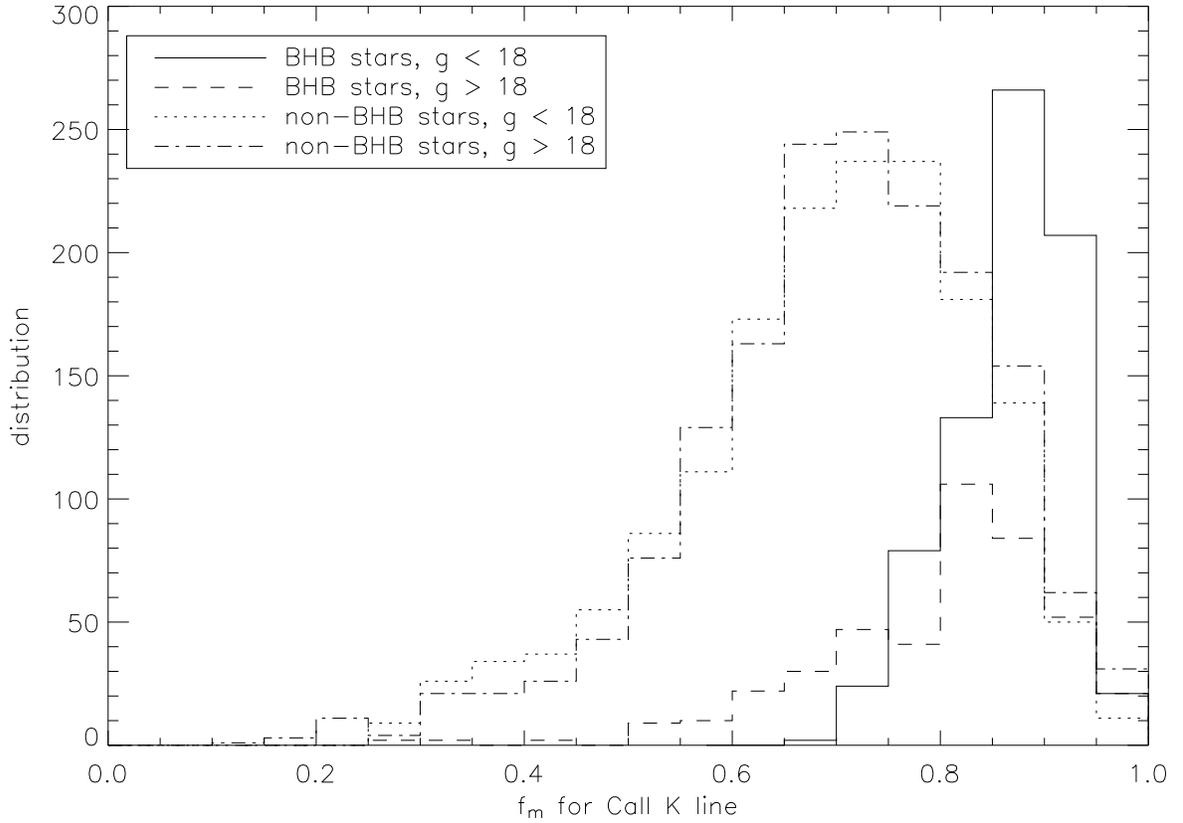}
\caption{Distribution of the $\fmx$ parameter for the CaII K line for BHB
stars and non-BHB stars.  Although BHB stars clearly have
larger values of $\fmx$ and thus lower metallicity implying that they are
halo objects, they are not so cleanly separated from the non-BHB stars.
This is probably because SDSS spectroscopy does not have high enough
resolution and S/N to probe the CaII K line as accurately as it can the Balmer
H$\gamma$ and H$\delta$ lines.  }
\label{f:caii}
\end{figure}

BHB stars are halo stars, and as such they should have lower
metallicity than the disk star contaminants.  The most promising
spectral feature to use for a metallicity determination is the CaII K
line at $3933.7 \ang$ (the H line, at $3968.5 \ang$, is not used 
because it is too close to H$\epsilon$).
Unfortunately, it is impractical to use the CaII K line as a \emph{reliable}
indicator of metallicity in individual SDSS spectra because much
higher spectral resolution would be necessary.  However, the following
exercise is executed anyway as a sanity check.  By employing a
continuum-subtraction algorithm very similar to that described in
\S\ref{s:d02_method} for the H$\gamma$ and H$\delta$ Balmer lines, we
find the value of $\fmx$ for the CaII K line; smaller values indicate
larger metallicity.  Figure~\ref{f:caii} presents the distributions of
this parameter for our sample of BHB stars and non-BHB stars passing
the color cut.  The figure shows different distributions for bright
($g < 18$) and faint ($g > 18$) stars because of the difference in S/N.  
Reassuringly, our BHB stars tend to have lower metallicities.

\subsection{Proper motion}\label{s:proper_motion}
\begin{figure}
\includegraphics[width=\textwidth]{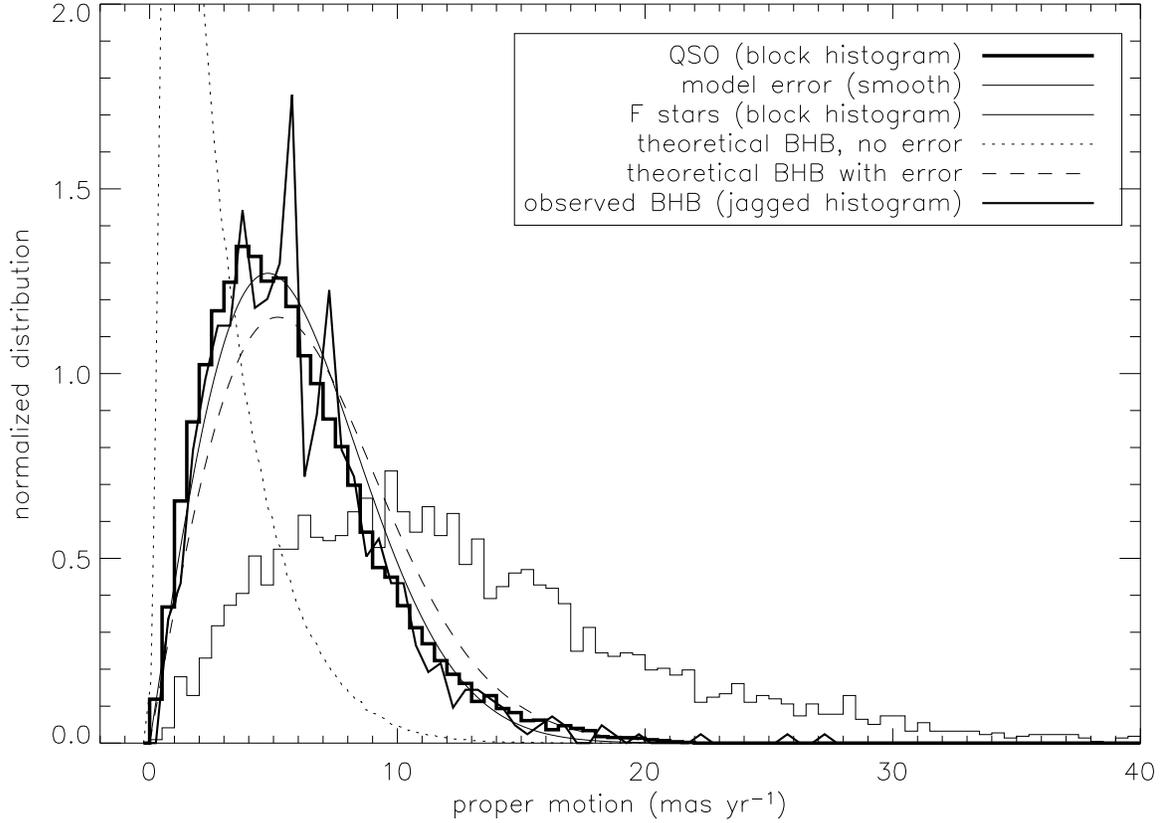}
\caption{Distributions of measured proper motion of SDSS objects, 
with integrals normalized
to unity, of SDSS objects compared to USNO catalogs.  
The QSO histogram reflects measurement
error.  The smooth thin curve is the expected analytic form
from equation~\protect{\ref{e:proper_motion}} of
the QSO distribution with variance $(4.77 \masyr)^2$.
Our sample of $\nbhb$ BHB stars is plotted as jagged (not boxy) for clarity.
The dotted line is the expected distribution of true proper motions for the
BHB sample.  The dashed curve is the
expected \emph{measured} distribution of BHBs given the
theory of measurement errors described in the text.  The close
match between the QSOs and BHB stars qualitatively 
indicates that contamination from foreground objects is low.}
\label{f:pm_dist}
\end{figure}

The proper motions of objects in SDSS 
are found automatically by matching the stellar positions to those in
the astrometric catalogue USNO-A V2.0 \citep{monet_etal_1998}
derived from the digitized Palomar Observatory Sky Survey (POSS) plates,
for objects bright enough to have a POSS counterpart. 
The normalized distribution of the proper motion of spectroscopically 
confirmed QSOs is shown in 
figure~\ref{f:pm_dist} as a thick line.  These alleged 
proper motions of QSOs are obviously due to measurement errors, 
both from POSS and SDSS (primarily from the former).  
If both measurements 
follow bivariate gaussian errors with radially symmetric 
variances $\sigma_1^2$ and 
$\sigma_2^2$, then the distribution of the separation of the measurements
(i.e., proper motion) also follows a bivariate gaussian with variance
$\sigma^2 = \sigma_1^2 + \sigma_2^2$.  Thus, the probability 
density distribution
of observing a star with true proper motion $\mu_0$ at an observed
proper motion $\mu$ is 
\begin{equation}\label{e:proper_motion}
\phi_{\mu_0}(\mu) = \frac{1}{2 \pi \sigma^2} \int_0^{2\pi} 
	\exp{\left(-\frac{\mu^2 + \mu_0^2 - 2\mu \mu_0 \cos{\theta}}
		{2\sigma^2}\right)} \mu \, d\theta = 
\frac{\mu}{\sigma^2} \exp{\left(-\frac{\mu^2 + \mu_0^2}{2\sigma^2}\right)}
	I_0 \left(\frac{\mu \mu_0}{\sigma^2} \right)
\end{equation}
where $I_0(x)$ is the modified Bessel function of the first kind.
Since $\mu_0 \to 0$ for the QSO distribution, the second moment of
the distribution of $\mu$ should equal $2 \sigma^2$ for gaussian
statistics.  This determination yields $\sigma_{\rm QSO} = 4.77 \masyr$, 
and the resulting theoretical
error distribution function from equation~\ref{e:proper_motion}
is shown in figure~\ref{f:pm_dist}
as a thin line.

We have estimated the statistical intrinsic proper motions of our sample using
the derived photometric distances.
The transverse velocity of each star was drawn 1000 times at random from
an isotropic velocity
distribution with dispersion $100\kms$ in one dimension and
corrected for an assumed solar velocity
$\vsun = (10, 225, 7) \kms$ \citep{dehnen_binney_1998}.
These choices were based on the kinematic results presented in 
Paper~II.
The distribution of intrinsic proper motions so derived,
$N(\mu_0)$, is shown
in figure~\ref{f:pm_dist} as the dotted line.  
The theoretical proper motion distribution for BHB stars with measurement
errors should follow the form
\begin{equation}\label{e:bhb_dist}
\phi_{\rm{BHB}}(\mu) \propto \int_0^\infty N(\mu_0) \phi_{\mu_0}(\mu) \, d\mu_0
\end{equation}
and is drawn in figure~\ref{f:pm_dist} as the dashed line.
The distribution of \emph{measured} proper motions of our 
BHB sample is also shown.
This distribution is 
clearly not very much broader than that of the QSOs; the second moment
yields $\sigma_{\rm BHB} = 4.84 \masyr$, which is actually
too small to account for the combination of measurement error 
($\sigma_{\rm QSO} = 4.77 \masyr$)
and intrinsic proper motion ($\sigma_{\rm BHB\,int} = 2.32 \masyr$)!  
In contrast, the proper motion distribution of 
a sample of faint (here $r > 16.7$) F stars from SDSS is also shown, 
with a much broader distribution.
This finding demonstrates that contamination in the BHB sample from F stars 
must be very small.  We estimate that the F star contamination is
less than $5\%$ among the 837 stars in our sample that were matched
to USNO-A.  Without better understanding of systematic
errors, we cannot claim to
have detected proper motion of the BHB sample.  However, the USNO-B
catalog, which has appreciably better data quality than USNO-A, will soon be
provided as the first epoch for proper motions, so we defer further 
analysis to later work.

\subsection{A map of the Galactic halo and the 
Sagittarius dwarf tidal stream}\label{s:sgr} 
\begin{figure}
\includegraphics[width=\textwidth]{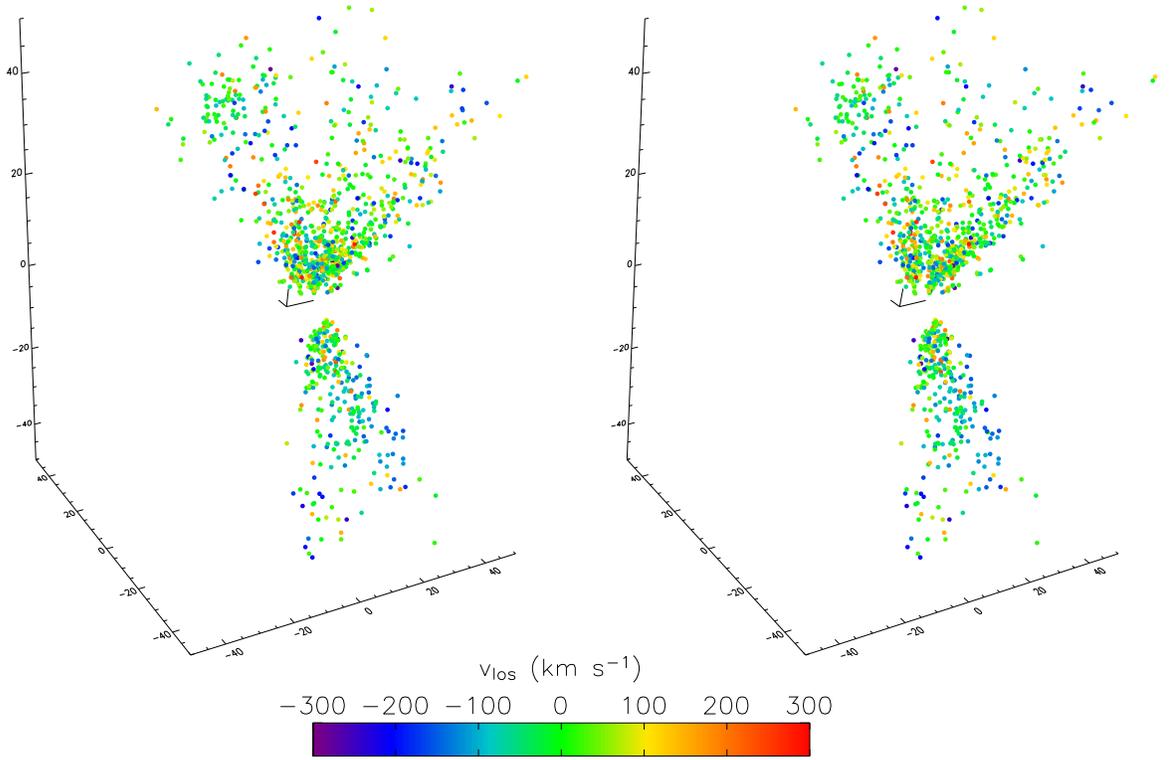}
\caption{Two three-dimensional projections of BHB stars discovered by the
SDSS spectroscopic survey, comprising a stereograph which can be 
viewed in 3-D by focusing the eyes on a point midway to the paper
or with a stereoscope.  Each star is colored
by its line-of-sight radial velocity with the solar velocity
subtracted out.
One can search for 
substructure in the halo by looking for streamers in this figure.
Although substructure is not the emphasis for this work, we note
the lump in the upper left which is the Sagittarius dwarf tidal stream.
The bracket in the middle of the figure represents a 
coordinate system centered on the Galactic center with the longest 
line segment pointing towards and terminating at the location of the sun
at $8 \kpc$.
Galactic north is up.  Only stars within a cube of half-length $50 \kpc$
are drawn.  The axes are labelled in kpc.}
\label{f:map_3d}
\end{figure}

Figure~\ref{f:map_3d} shows two views of a three-dimensional
representation of the locations of the BHB stars. Each star is
colored according to its line-of-sight radial velocity with
the solar velocity of
$(10, 225, 7) \kms$ \citep{dehnen_binney_1998} added.  These two views
compose a stereograph which can be viewed by focusing the eyes on
a point midway to the paper.  The
three-dimensional positions of the stars are derived from their
angular positions and photometric distances (\S\ref{s:absmag}), assuming
$R_0=8 \kpc$.

\begin{figure}
\includegraphics[width=\textwidth]{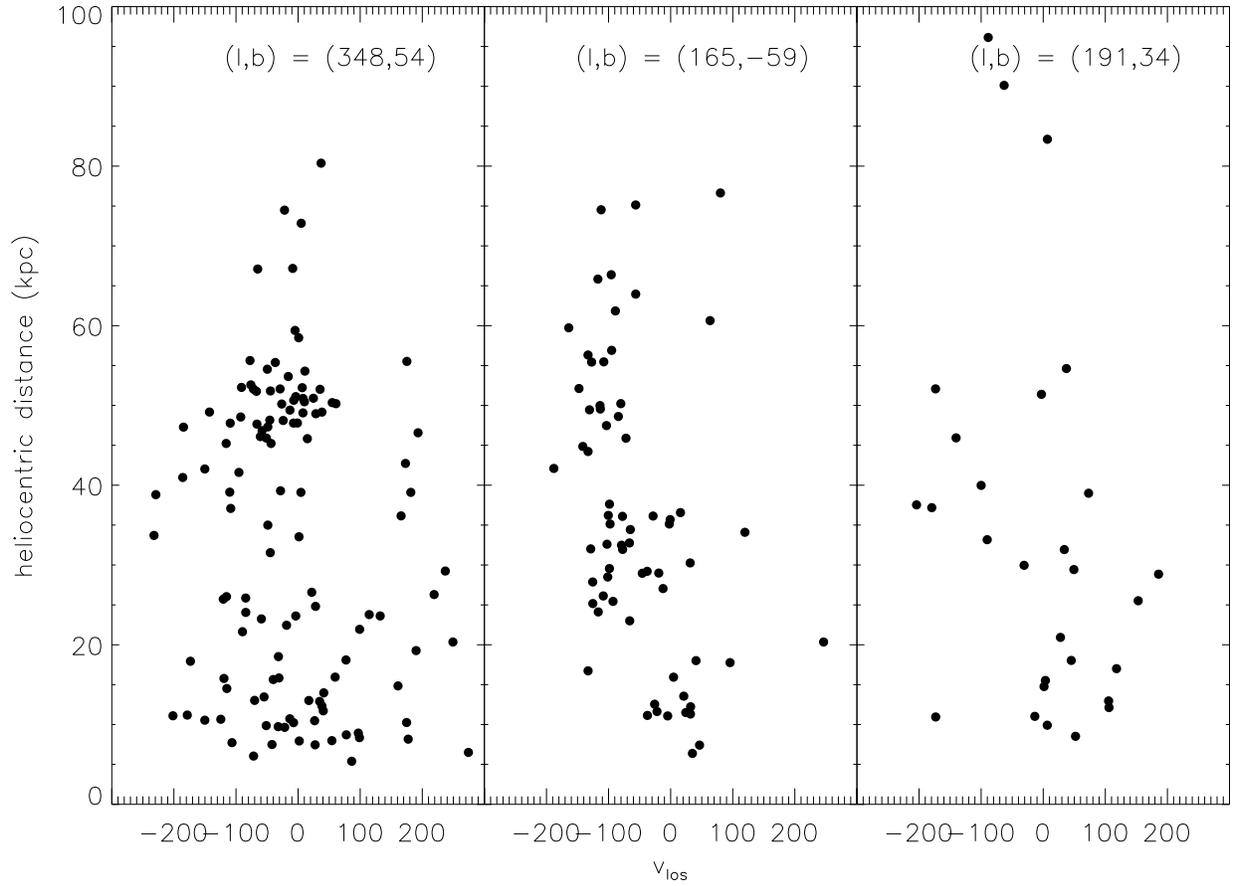}
\caption{Phase space plots for three subsamples of the BHB sample extracted
from the Sagittarius stream.  The three regions of extraction are given
by the intersection of the BHB sample with the
10 degree wide strip following the great 
circle described in \S\ref{s:sgr}.  The left plot,
from the north Galactic cap, clearly shows a clump in phase space 
at $\sim 50 \kpc$ which
represents the Sagittarius stream.  The middle plot, from the south,
can also be seen to show similar structure.  The right plot is from
the north but the number of stars is too small to make any
conclusions about the detection of the Sagittarius stream in this region
of space.}
\label{f:sgr_d_v_v}
\end{figure}

One could search for substructure in the Galactic halo by a careful
examination of figure~\ref{f:map_3d}, and indeed a clump appears in
the upper left part of the plot (at $(\ell,b) \sim (350,50)$).  
This clump belongs to the leading arm of the Sagittarius dwarf tidal 
stream \citep{ivezic_etal_2000,yanny_etal_2000,
ibata_etal_2001a,ibata_etal_2001b}.  
It is interesting
to investigate this structure in velocity phase space, as shown in
figure~\ref{f:sgr_d_v_v}.  
The Sagittarius stream approximately follows the great circle on the celestial 
sphere with pole $(\alpha,\delta) = (308,58)$ or $(\ell,b) = (94,11)$, as
determined from the distribution of SDSS candidate RR Lyrae stars. Practically
the same position is implied by the distribution of 2MASS M giants 
\citep{majewski_etal_2003}.
This great circle
intersects the footprint of the SDSS in three places.  These three
intersections with BHB stars (two in the north and one in the south,
with a strip width of 10 degrees) are represented in the three panels
in figure~\ref{f:sgr_d_v_v}.  One can see a clump in the phase space
of the northern subsample representing the Sagittarius stream (left
plot); indeed it is the same clump in the upper left of
figure~\ref{f:map_3d}.  
The coherent structure in the middle panel (at $D \sim 30 \kpc$) is
also associated with the Sgr dwarf tidal stream (trailing arm). 
This is the same clump that was discovered by \citet{yanny_etal_2000}
using A-colored stars, and is also detected in the 
velocity-distance distribution of SDSS candidate RR Lyrae stars 
\citep{ivezic_etal_2003}.  The fact that we see such structure
in the BHB sample strongly indicates that our BHB absolute magnitudes
and selection criteria are reliable.

\section{Summary}\label{s:summary}

Using broadband colors and spectroscopic surface-gravity indicators,
we have selected $\nbhb$ blue horizontal branch stars
from the SDSS.  For the $\nbhbbr$ bright stars ($g<18$) the contamination
is probably $\lesssim 10\%$, and for the $\nbhbfaint$ faint BHB
stars ($g>18$) the contamination is probably $\sim 25\%$.
The absolute magnitude of BHB stars was determined
from globular cluster color-magnitude diagrams, and an 
absolute-magnitude-color relation established. This was used to
estimate photometric distances, with errors $\approx 0.2$ mag. 
The stars also have measured
radial velocities with errors $\approx 26 \kms$.
Proper motions determined by matching against the
digitally scanned Palomar Sky Survey \citep[USNO-A,][]{monet_etal_1998}
also indicate that contamination from the disk is small.
We emphasize, however, that no kinematic information was used to
select the sample.
The data can be downloaded as Table~\ref{t:data} in the
electronic version of this paper.

\acknowledgements
We thank A.~Gould, B.~Paczy\'nski, R.~Lupton, and
J.~Hennawi for helpful discussions, and M.~Strauss for a close reading
of a draft of this paper.
GRK is grateful
for generous research support from Princeton University and
from NASA via
grants NAG-6734 and NAG5-8083.
Funding for the creation and distribution of the SDSS Archive has been
provided by the Alfred P.
Sloan Foundation, the Participating Institutions, the National Aeronautics
and Space Administration, the National Science Foundation, the U.S.
Department of Energy, the Japanese Monbukagakusho, and the Max Planck Society.  The SDSS Web
site is {\tt http://www.sdss.org/}.  The SDSS is managed by the
Astrophysical Research Consortium (ARC) for the Participating Institutions.
The Participating Institutions are The University of Chicago, Fermilab,
the Institute for Advanced Study, the Japan Participation Group, The Johns
Hopkins University, Los Alamos National Laboratory, the Max-Planck-Institute
for Astronomy (MPIA), the Max-Planck-Institute for Astrophysics (MPA),
New Mexico State University, the University of Pittsburgh, Princeton
University, the United States Naval Observatory, and the University of
Washington.

\appendix
\section{Remark on contamination and incompleteness}\label{s:contamination}
There is no independently confirmed
subsample of BHB stars, but we can estimate the accuracy of our
spectroscopic selection methods by intercomparing their results.

After applying our color cut, we are left with $N+C=\ncolorcutbr$ candidates
bright enough to be subjected to our spectroscopic selection criteria. Of
these, $N$ are true BHB stars, and the other $C$ are potential contaminants.
Let $\eta_i\le1$ be the probability that method $i$ fails to identify
a true BHB as such, and $\kappa'_i$ be the probability that it mistakenly
accepts an individual member of the non-BHB group.  Then the
expected number of stars accepted by method $i$ is
\begin{equation}\label{e:Ni_expected}
N_i= (1-\eta_i)N ~+\kappa'_i C
\end{equation}
Similarly, the numbers of stars selected jointly by methods $i$ \& $j$, or
by $i$, $j$, \& $k$ , are
\begin{eqnarray*}
N_{i\cap j}&=& (1-\eta_i)(1-\eta_j)N~+\kappa'_i\kappa'_j C\,,\\
N_{i\cap j\cap k}&=& (1-\eta_i)(1-\eta_j)(1-\eta_k)N~
+\kappa'_i\kappa'_j\kappa'_k C\,,
\end{eqnarray*}
respectively, assuming that these methods are truly independent.
It is convenient to define $\kappa_i\equiv\kappa'_i C/N_i$ so that
the number of contaminants among the $N_i$ stars accepted by method $i$ is
$\kappa_i N_i$.  Then
\begin{eqnarray}
\label{e:Ni_eqn}
N_i&=&\frac{(1-\eta_i)}{(1-\kappa_i)}N,\\
\label{e:2ndorder}
N_{i\cap j}&=& \left[\frac{(1-\kappa_i)(1-\kappa_j)}{N}
~+\frac{\kappa_i\kappa_j}{C}\right]N_iN_j \\
\label{e:3rdorder}
N_{i\cap j\cap k}&=& \left[\frac{(1-\kappa_i)(1-\kappa_j)(1-\kappa_k)}{N^2}
~+\frac{\kappa_i\kappa_j\kappa_k}{C^2}\right]N_iN_jN_k
\end{eqnarray}

As described in \S\ref{s:balmer_cut},
we have four methods.  Therefore we have six
double intersections (\ref{e:2ndorder}) and four
triple intersections (\ref{e:3rdorder}) with which to solve for the
six unknown parameters $(\kappa_1,\ldots\kappa_4,N,C)$ that they contain.  So
the problem is overdetermined, allowing us perhaps to test the assumption
of independence. Perfect consistency is not to be expected
since even for fixed values of the above parameters, the numbers
$\{N_i,N_{i\cap j},N_{i\cap j\cap k}\}$ are subject to 
statistical fluctuations.

If the contamination fractions $\kappa_i$ are reasonably small---it is
sufficient that
$\kappa/(1-\kappa)\ll\sqrt{C/N}$---then the terms involving $C$ in
eqs.~(\ref{e:2ndorder})-(\ref{e:3rdorder}) are unimportant, so that
\begin{equation}\label{e:Nratios}
1-\kappa_i \approx \frac{N_{i\cap j}N_{i\cap k}}{N_{i\cap j\cap k}N_i}
\end{equation}
This approach yields the results shown in Table~\ref{t:accuracy}, and
also an estimate for the true number of BHBs tested (\emph{i.e}, not
including those too faint for spectroscopic selection): $N\approx 772$.  
Note that there are three possible choices of $(j,k)$ with which to
estimate $\kappa_i$ from eq.~(\ref{e:Nratios}), all of which are shown
in the Table to give an idea of their consistency.
It can
be seen that the contamination and incompleteness fractions for each are
indeed small, $<10\%$, if these methods are indeed independent as we have
assumed.  In support of the latter,
the differences among the estimates in columns
2-4 are $\sim\pm0.01$ as would be expected by chance since, for example,
binomial statistics predict that
the number of false negatives for method $i$ should fluctuate by
$\pm\sqrt{\eta_i(1-\eta_i)N}$, or about 7 out of 772 if $\eta_i=0.07$.

\clearpage

% --- Bibliography  -----------------------------------

\clearpage

% --- Begin figures -----------------------------------
%f:ugr_g18: ug-gr diagram with box showing color cut (stars_about).

%f:bhb_spectrum. (plotspec_paper,301,420).

%f:d02_demo (d02_demo,290,301,366,132).

%f:d02_v_fm (d02_v_fm_and_ava)

%f:g18_breakdown (g18_breakdown).

%f:compare1 (stars_lw_sersic).

%f:compare_rhb (compare_rhb).

%f:ugr_bhb18: ug-gr diagram, zoomed, with BHB indicated (stars_about).

%f:duplicates (more_duplicates_analysis)

%f:gc_cmd (analyze_gchb)

% do this to avoid "too many unprocessed floats"
\clearpage

%f:caii (sanity_checks)

%f:pm_dist (pm_theory_dist,pm_convolve)

%f:map_3d (stars_map)

%f:sgr_d_v_v (sgr_d_v_v)

\clearpage
% --- Begin tables  -----------------------------------

%t:criteria
\begin{deluxetable}{lll}
\tablewidth{0pt}
\tablecolumns{3}
\tablecaption{BHB criteria for the different methods \label{t:criteria}}
\tablehead{ \colhead{method} & \colhead{temperature criterion} &
    \colhead{linewidth criterion}}
\startdata
$D_{0.2}$ method (H$\gamma$) & $\fmx < 0.37$ & $D_{0.2} < 28\ang$ \\
$D_{0.2}$ method (H$\delta$) & $\fmx < 0.34$ & $D_{0.2} < 28\ang$ \\
scalewidth-shape method (H$\gamma$) & $c > 1.00$ & $b < 12.0 - 30(c-1.18)^2$ \\
scalewidth-shape method (H$\delta$) & $c > 0.95$ & $b < 12.5 - 40(c-1.10)^2$ \\
\enddata
\end{deluxetable}

%t:thb
\begin{deluxetable}{cccccc}
\tablecaption{Predicted absolute magnitudes and colors for blue horizontal
branch stars in the SDSS system, assuming
eq.~(\ref{e:MLrels}) and $[M/H]=-2.0$ (1st line at each $\Teff$) and
$[M/H]=-1.0$ (2nd line). In the second column $g$ is gravity; in
all other instances it is the magnitude in the $g$ band.
	\label{t:thb}}
\tablewidth{0pt}
\tablehead{\colhead{$\Teff$}&\colhead{$\log g$}&\colhead{$g(10\pc)$}&
	\colhead{$u-g$}&\colhead{$g-r$}&\colhead{$g-i$}
}
\startdata
  7000 & 2.900  &  0.7115 &  1.0335 &  0.0945 &  0.0736 \\
       &        &  0.6862 &  1.0836 &  0.0999 &  0.0724 \\ [1.0ex]
  8000 & 3.132  &  0.5712 &  1.1792 & -0.1313 & -0.2700 \\
      &         &  0.5494 &  1.1944 & -0.1291 & -0.2724 \\ [1.0ex]
  9000 & 3.336  &  0.6311 &  1.0695 & -0.2300 & -0.4301 \\
      &         &  0.6121 &  1.0732 & -0.2317 & -0.4351 \\ [1.0ex]
 10000 & 3.520  &  0.8000 &  0.8911 & -0.2742 & -0.5023 \\
      &         &  0.7753 &  0.8853 & -0.2780 & -0.5093 \\ [1.0ex]
 11000 & 3.685  &  0.9954 &  0.7381 & -0.2999 & -0.5449 \\
      &         &  0.9672 &  0.7239 & -0.3034 & -0.5512 \\ [1.0ex]
 12000 & 3.836  &  1.1884 &  0.6139 & -0.3203 & -0.5785 \\
      &         &  1.1582 &  0.5950 & -0.3232 & -0.5836 \\ [1.0ex]
 13000 & 3.975  &  1.3735 &  0.5103 & -0.3382 & -0.6075 \\
      &         &  1.3425 &  0.4881 & -0.3404 & -0.6115 \\
\enddata
\end{deluxetable}

%t:data
\begin{deluxetable}{llllclllllclllllclllcllllclcllll} % 26+6 of them
\tablecaption{List of $\nbhb$ BHB stars selected from SDSS\label{t:data}}
\rotate
\tabletypesize{\scriptsize}
\tablewidth{0pt}
\tablehead{
\multicolumn{4}{c}{astrometry} & \colhead{} &
\multicolumn{5}{c}{uncorrected magnitudes} & \colhead{} &
\multicolumn{5}{c}{extinction-corrected magnitudes} & \colhead{} &
\multicolumn{3}{c}{} & \colhead{} &
\multicolumn{4}{c}{$D_{0.2}$ method parameters} & \colhead{} &
\colhead{} & \colhead{} &
\multicolumn{4}{c}{scalewidth-shape parameters} \\
\cline{1-4} \cline{6-10} \cline{12-16} \cline{22-25} \cline{29-32} \\
\colhead{ra} & \colhead{dec} & \colhead{$\ell$} & \colhead{$b$} & \colhead{} &
\colhead{$u$} & \colhead{$g$} & \colhead{$r$} & 
	\colhead{$i$} & \colhead{$z$} & \colhead{} &
\colhead{$u$} & \colhead{$g$} & \colhead{$r$} & 
	\colhead{$i$} & \colhead{$z$} & \colhead{} &
\colhead{redshift} & \colhead{$\vlos$} & 
	\colhead{$D~(\rm{kpc})$} & \colhead{} &
\colhead{$D_{0.2,\rm{H}\gamma}$} & \colhead{$f_{m,\rm{H}\gamma}$} & 
	\colhead{$D_{0.2,\rm{H}\delta}$} & \colhead{$f_{m,\rm{H}\delta}$} & 
	\colhead{} &
\colhead{$f_{m,\rm{CaII}}$} & \colhead{} &
\colhead{$b_{\rm{H}\gamma}$} & \colhead{$c_{\rm{H}\gamma}$} & 
	\colhead{$b_{\rm{H}\delta}$} & \colhead{$c_{\rm{H}\delta}$}
}
\startdata
  145.82320 &   -1.17255 &  237.13901 &   36.66249 & & 17.89 & 16.68 & 16.85 & 16.95 & 17.04 & & 17.65 & 16.50 & 16.73 & 16.86 & 16.97 & &  0.0005883 &   24.71
&  15.34 & & 27.3646 &  0.3316 & 25.6610 &  0.2890 & &  0.9099 & & 11.5214 &  1.1931 & 11.2208 &  1.1947 \\
  160.36295 &   -1.05517 &  249.66811 &   47.91172 & & 18.08 & 16.83 & 16.91 & 16.94 & 17.06 & & 17.80 & 16.62 & 16.76 & 16.83 & 16.98 & &  0.0005751 &   33.98
&  16.36 & & 25.9615 &  0.3286 & 28.0140 &  0.2953 & &  0.8329 & & 11.0010 &  1.1813 & 11.9872 &  1.0941 \\
  161.66508 &    0.06255 &  249.84621 &   49.62371 & & 18.55 & 17.20 & 17.36 & 17.47 & 17.50 & & 18.34 & 17.05 & 17.25 & 17.39 & 17.44 & &  0.0005553 &   32.86
&  19.98 & & 24.1510 &  0.3700 & 27.3331 &  0.3313 & &  0.8483 & & 10.4411 &  1.0110 & 10.8343 &  0.9828 \\
  161.08659 &    1.00213 &  248.17850 &   49.86035 & & 18.64 & 17.37 & 17.40 & 17.51 & 17.55 & & 18.41 & 17.20 & 17.28 & 17.42 & 17.48 & &  0.0008044 &  109.62
&  21.27 & & 23.5976 &  0.3580 & 27.3562 &  0.3028 & &  0.8724 & &  9.3291 &  1.0594 & 11.6956 &  0.9662 \\
  161.20377 &    0.68167 &  248.66331 &   49.72445 & & 17.88 & 16.60 & 16.72 & 16.81 & 16.89 & & 17.65 & 16.43 & 16.60 & 16.72 & 16.82 & &  0.0006674 &   67.72
&  14.98 & & 26.9836 &  0.3368 & 26.8800 &  0.3101 & &  0.7589 & & 10.5769 &  1.0842 & 11.7013 &  1.0022 \\
\enddata
\tablecomments{The first four columns contain the astrometry 
(ra, dec, $\ell, b$) for each object.  
The magnitudes $(u,g,r,i,z)$ are in the next 
ten columns: uncorrected and corrected for extinction.  The heliocentric
redshift is listed next.  The next column is $\vlos$, the
galactocentric velocity in $\kms$, assuming $\vsun = (10,5,7) \kms$.  
The next column is the distance in kpc
as determined by the method of \S\protect{\ref{s:absmag}}.  The last 
nine columns are the linewidth parameters from the Balmer and CaII lines.
The complete version of this table is in the 
electronic edition of the Journal.  
The printed edition contains only a sample.}
\end{deluxetable}

% Tables associated with appendix.  I suggest that these two replace the
% existing tables 2-4. ---Jeremy
\begin{deluxetable}{ccccc}
\tablewidth{0pt}
\tablecaption{Numbers of stars jointly selected\label{t:jointsel}}
\tablehead{\colhead{$D_{0.2}(\rm{H}\gamma)$} & 
	\colhead{$D_{0.2}(\rm{H}\delta)$} &
	\colhead{SW-S (H$\gamma$)} &
	\colhead{SW-S (H$\delta$)} &
	\colhead{$N_{i}, N_{i\cap j} \cdots$} }
\startdata
 0 &  0 &  0 &  0 &  2338 \\
 0 &  0 &  0 &  1 &   741 \\
 0 &  0 &  1 &  0 &   746 \\
 0 &  0 &  1 &  1 &   683 \\
 0 &  1 &  0 &  0 &   779 \\
 0 &  1 &  0 &  1 &   661 \\
 0 &  1 &  1 &  0 &   658 \\
 0 &  1 &  1 &  1 &   620 \\
 1 &  0 &  0 &  0 &   777 \\
 1 &  0 &  0 &  1 &   670 \\
 1 &  0 &  1 &  0 &   692 \\
 1 &  0 &  1 &  1 &   642 \\
 1 &  1 &  0 &  0 &   673 \\
 1 &  1 &  0 &  1 &   620 \\
 1 &  1 &  1 &  0 &   630 \\
 1 &  1 &  1 &  1 &   595 \\
\enddata
\tablecomments{
The first four columns are binary bits indicating which selection 
methods were applied.
The top line is the total number of candidate stars ($g<18$);
the bottom line is the number selected by all four methods.}
\end{deluxetable}

\begin{deluxetable}{lllllll}
\tablewidth{0pt}
\tablecaption{Accuracy of four spectroscopic selection methods.
\label{t:accuracy}}
\tablehead{
\colhead{method $i$} & \multicolumn{3}{c}{3 estimates of $\kappa_i$} & 
\colhead{$\langle\kappa_i\rangle$} & \colhead{$\langle\eta_i\rangle$}}
\startdata
$D_{0.2}(\rm{H}\gamma)$ & 0.0706 & 0.0640 & 0.0486      & 0.0611    & 0.0552 \\
$D_{0.2}(\rm{H}\delta)$ & 0.0995 & 0.0789 & 0.0977      & 0.0920    & 0.0840 \\
$\mbox{SW-S}(\rm{H}\gamma)$ & 0.0283 & 0.0131 & 0.0312  & 0.0242    & 0.0573 \\
$\mbox{SW-S}(\rm{H}\delta)$ & 0.0173 & 0.0381 & 0.0360  & 0.0305    & 0.0696 \\
\enddata
\tablecomments{Estimates in columns 2-4 are based on the three
triples involving $i$ drawn from methods $(1,2,3,4)$
[eq.~(\ref{e:Nratios})], column 5
is their average, and column 6 is computed from column 5
and eq.~(\ref{e:Ni_eqn}).}
\end{deluxetable}

\end{document}